\documentclass[aps,prd,showpacs,preprintnumbers,superscriptaddress,preprint,groupedaddress]{revtex4}
\usepackage{amsmath}
\usepackage{color}
\usepackage{graphicx}
\usepackage{bm}
\usepackage{hyperref}
\allowdisplaybreaks[3]

\usepackage[normalem]{ulem}  % \sout{old text} for strikeout
\renewcommand\sout{\bgroup \color{red} \ULdepth=-.5ex \ULset}

\begin{document}

\preprint{YITP-19-83}
\preprint{iTHEMS-Report-19}
\title{Hermitizing the HAL QCD potential in the derivative expansion}

%%%%%
\author{Sinya~Aoki}
\email{saoki@yukawa.kyoto-u.ac.jp }
\affiliation{Center for Gravitational Physics,
Yukawa Institute for Theoretical Physics, Kyoto University, Kyoto 606-8502, Japan}
%%%%%%
\author{Takumi~Iritani}
\email{takumi.iritani@alum.riken.jp}
\affiliation{Theoretical Research Division, Nishina Center, RIKEN, Saitama 351-0198, Japan}
\author{Koichi~Yazaki}
\email{koichiyzk@yahoo.co.jp}
\affiliation{Interdisciplinary Theoretical and Mathematical Sciences Program (iTHEMS), RIKEN Saitama 351-0198, Japan} 
%%%%%%

\date{\today}

\begin{abstract}
A formalism is given to hermitize the HAL  QCD potential, which needs to be  non-hermitian except  the leading order (LO) local term in the derivative expansion as the Nambu-Bethe-Salpeter (NBS) wave functions for different energies are not orthogonal to each other.
 It is shown that the non-hermitian potential can be hermitized order by order  to all orders in the derivative expansion. In particular,  the next-to-leading order (NLO) potential  can be exactly hermitized without approximation.   
 The formalism is then applied to a simple case of $\Xi \Xi (^{1}S_{0}) $ scattering, for which the HAL QCD calculation is available to the NLO. The NLO term gives relatively small corrections to the scattering phase shift and the LO analysis seems justified in this case.
We also observe that the local part of the hermitized NLO potential works better than that of the non-hermitian NLO potential. 
The hermitian version of the HAL QCD potential is desirable for comparing it with phenomenological interactions and also for using it as a two-body interaction in many body systems.

\end{abstract}


\maketitle

\section{Introduction}
\label{Introduction}
  Lattice quantum chromodynamics (QCD) is a successful non-perturbative method to study hadron physics from 
  the underlying degrees of freedom, i.e. quarks and gluons.
  Masses of the single stable hadrons obtained from lattice QCD show good agreement with the experimental results, 
  and even hadron interactions have been recently explored in lattice QCD.
  Using the Nambu-Bethe-Salpeter (NBS) wave function, linked to the S-matrix in QCD
   \cite{Luscher:1990ux, Lin:2001ek, Aoki:2005uf, Ishizuka:2009bx,
   Ishii:2006ec, Aoki:2009ji, Aoki:2012tk, 
Carbonell:2016ekx,Aoki:2013cra}, 
 the hadron interactions have been investigated mainly by two methods: the finite volume method \cite{Luscher:1990ux} and 
  the HAL QCD potential method \cite{ Ishii:2006ec,Aoki:2009ji,Aoki:2012tk}. Theoretically the two methods in principle give same
  results of the scattering phase shifts between two hadrons, while in practice they sometimes show different numerical
   results for two baryon systems, whose  origin has been clarified recently in Refs.~\cite{Iritani:2016jie,Iritani:2017rlk}.

  The HAL QCD method utilizes the NBS wave function in
  non-asymptotic (interacting) region, and extract the non-local but energy-independent potentials from the space and time dependences of the NBS wave function.
  Physical observables such as phase shifts and binding energies are then calculated by solving the Schr\"odinger equation in infinite volume using the obtained potentials,
  since  the asymptotic behavior of the NBS wave function is related to the $T$-matrix  element
and thus to the phase shifts \cite{Aoki:2013cra}. 
 In practice, the non-local potential is given by the form of the derivative expansion,
 which is truncated by the first few orders \cite{Iritani:2018zbt}.
 
 While the HAL QCD method has been successfully applied to a wide range of two (or three) hadron systems at heavy pion masses 
  \cite{Nemura:2008sp, Inoue:2010hs, Inoue:2010es, Murano:2011nz, Doi:2011gq,Inoue:2011ai,  Murano:2013xxa, Kurth:2013tua, Ikeda:2013vwa, Etminan:2014tya, Yamada:2015cra, Sasaki:2015ifa, Ikeda:2016zwx, Miyamoto:2017tjs, Kawai:2017goq,Ikeda:2017mee} 
  as well as at the nearly physical mass \cite{Gongyo:2017fjb,Sasaki:2017ysy,Ishii:2017xud, Doi:2017cfx, Nemura:2017bbw,Doi:2017zov,Nemura:2017vjc},
there are some subtleties or issues in the method.
One is  the theoretical treatment of the bound states in this method, which has been recently clarified
in Ref.~\cite{Gongyo:2018gou}. 
In this paper, we consider the other issue,  non-hermiticity of the potential
 in the HAL QCD method.
We show in Sec.~\ref{sec:hermitizing} that non-hermitian potential defined in the derivative expansion can be made hermitian order by order in the derivative expansion. 
In particular, non-hermitian potential which contains the second derivative at most can be made 
hermitian exactly, while
non-hermitian potentials with higher order derivative than the second order can be shown to be made hermitian potentials, using the mathematical induction for the order of the derivative expansion.
In Sec.~\ref{sec:application}, we apply our method to a non-hermitian HAL QCD potential 
for $\Xi\Xi$ in lattice QCD\cite{Iritani:2018zbt}, which consists of local and second or first order derivative terms.
We  show that the exactly hermitized potential gives the same scattering phase shifts with those from the original non-hermitian potential but the contribution from its derivative term is smaller than the one from the original derivative term.
The summary and conclusion of this paper is given  in Sec.~\ref{sec:summary}.

\section{Hermitizing the non-hermitian potential}
\label{sec:hermitizing}
In this section, we propose a method to hermitize the non-hermitian Hamiltonian order by order
in terms of derivatives.
We consider the non-hermitian Hamiltonian for the relative coordinate of two identical particles without spin, which is given by
\begin{eqnarray}
H &=& H_0 + \sum_{n=0}^\infty V_n,  \quad H_0 = -\frac{1}{m}\nabla^2, 
\end{eqnarray}
where $V_n$ is the potential with $n$-derivatives, and $m$ is the mass of the particle, so that the reduced mass is given by $m/2$.
The explicit form of $V_n$ is denoted as
\begin{eqnarray}
V_n &:=&\frac{1}{n !} V_l^{i_1i_2\cdots  i_n}\nabla_{i_1}\nabla_{i_2}\cdots \nabla_{i_n},
\label{eq:V_n}
\end{eqnarray}
where the local function $V_l^{i_1i_2\cdots  i_n}$ is symmetric under exchanges of indices $i_1i_2\cdots  i_n$ and summations over repeated indices are implicitly assumed.
In this paper, we assume that the above derivative expansion is convergent.
See appendix~\ref{app:conv} for some arguments.
Except the local potential $V_0$, other $V_{n> 0}$ are non-hermitian.
Note also that the $\vec r$-dependence of $V_n$ is also implicit.

Since $H_0+V_0$ is hermitian, we first consider  $V_1$ and $V_2$, which is  the next-to-leading (1st) order, and
more generally $V_{2n-1}$ and $V_{2n}$ as the $n$-th order,
for the hermitizing problem, 
and introduce
\begin{eqnarray}
U_n &:=& V_{2n} + V_{2n-1}.
\label{eq:U_n}
\end{eqnarray}
The reason to treat $V_{2n}$ and $V_{2n-1}$ together will be clear later.
In terms of the derivative expansion for the potential, 
$V_0$ is of leading order while $V_1$ and $V_2$ are of next-to-leading, so that $V_0$ is much larger in size than $V_1$ or $V_2$ at low energies.

\subsection{$n=1$ case}
At $n=1$, the Hamiltonian is given by
\begin{eqnarray}
H^{(1)} = H_0 + V_0 + U_1,
\end{eqnarray}
where the $n=1$ potential $U_1$ is rewritten as
\begin{eqnarray}
U_1 &=& \tilde V_2 + \tilde V_1, \quad  \tilde V_2: =\frac{1}{2} \nabla_i V_2 ^{ij}\nabla_j,
\quad \tilde V_1 := \tilde V_1^i \nabla_i, \quad
\tilde V_1^i :=  V_1^i -\frac{1}{2}\left( \nabla_j V_2^{ji} \right).
\end{eqnarray}
Here $\tilde V_2$ is hermitian, while $\tilde V_1$ is not.

The corresponding Schr\"odinger equation is given by
\begin{eqnarray}
H^{(1)} \psi &=& E\psi,
\end{eqnarray}
which transforms to
\begin{eqnarray}
\tilde H^{(1)} \phi = E\phi, \qquad \tilde H^{(1)} = R_1^{-1} H^{(1)} R_1,
\end{eqnarray}
by the change of the wave function that $\psi := R^{(1)} \phi$ with a local function $R^{(1)}=R_1$, 
where
\begin{eqnarray}
\tilde H^{(1)} &=& H_0 +\tilde V_0 +\tilde V_2 +\left\{ \tilde V_1^i -\frac{2}{m} R_1^{-1} \nabla^i R_1 + V_2^{ij} R_1^{-1} \nabla_j R_1 \right\}\nabla_i, 
\\
\tilde V_0 &=& V_0 -\frac{1}{m} R_1^{-1} \nabla^2 R_1 + V_1^i R_1^{-1}\nabla_i R_1 
+\frac{1}{2} V_2^{ij}\left(R_1^{-1}\nabla_i\nabla_j R_1\right)
.
\end{eqnarray}
By demanding the condition that
\begin{eqnarray}
\tilde V_1^i -\frac{2}{m} R_1^{-1} \nabla^i R_1 + V_2^{ij} R_1^{-1} \nabla_j R_1&=& 0,
\end{eqnarray}
$\tilde H^{(1)}$ becomes hermitian as  $\tilde H^{(1)} = H_0 +\tilde V_0 +\tilde V_2$, where
\begin{eqnarray}
\tilde V_0 &=& V_0 -\frac{1}{2}\left( \nabla_i \tilde V_1^i \right) +\frac{m}{4} \tilde V_1^i
\left(\delta^{ij} - \frac{m}{2} V_2^{ij} \right)^{-1} \tilde V_1^j , \\
R_1^{-1} \nabla_i R_1 &=&\frac{m}{2} \left(\delta^{ij} - \frac{m}{2} V_2^{ij} \right)^{-1} \tilde V_1^j \ . 
\end{eqnarray}

In the rotationally symmetric case such that 
\begin{eqnarray}
\tilde V_1^i(\vec r) & :=& \hat {r^i} \tilde V_1 (r), \quad  V_2^{ij}(\vec r) :=
 V_{2a}(r) \hat r^i \hat r^j+ V_{2b}(r) \delta^{ij},  \quad  R_1(\vec r) := R_1(r)
\end{eqnarray}
with $r:=\vert\vec r\vert$ and $\hat r^i := r^i/r$, we have
\begin{eqnarray}
\frac{d\, R_1(r)}{d\, r} &=& \frac{m}{2} \frac{\tilde V_1(r)}{1-\dfrac{m}{2} V_2(r)} R_1(r) ,
\qquad
V_2:= V_{2a}+V_{2b}, 
\end{eqnarray}
which can be solved as
\begin{eqnarray} 
R_1(r) &=& \exp\left[ \frac{m}{2}\int_{r_\infty}^r \frac{\tilde V_1(s)}{1-\dfrac{m}{2} V_2(s)}\, d\, s \right], 
\end{eqnarray}
where we assume $V_1(r) =0$ and $R_1(r)=1$ at sufficiently large $r\ge r_\infty$.
Thus the hermitian local potential $\tilde V_0$ becomes 
\begin{eqnarray}
\tilde V_0 &=& V_0 -\frac{\tilde V_1}{r} -\frac{1}{2} \tilde V_1^\prime + \frac{m}{4} \frac{\tilde V_1^2}{1-\dfrac{m}{2} V_2}, \qquad
\tilde V_1 := V_1 - \frac{V_{2a}}{r} -\frac{V_{2a}^\prime + V_{2b}^\prime}{2},
\end{eqnarray}
where the prime $^\prime$ means the derivative with respect to $r$.

\subsection{$n=2$ case}
In the previous subsection, we show that the non-hermitian potential at $n=1$
can be made hermitian without any approximations. In this subsection, we proceed to the next order, the $n=2$ case,
 where some truncations are required for a number of derivatives, as we will see. 

The $n=2$ Hamiltonian is given by
\begin{eqnarray}
H^{(2)} = H^{(1)}+ U_2,
\end{eqnarray}
where the $n=2$ potential $U_2$ can be written as
\begin{eqnarray}
U_2 &=& \frac{1}{4!} \nabla_i\nabla_j\, U_{2,4}^{ijkl}\nabla_k\nabla_l +
 \frac{1}{3!}  U_{2,3}^{ijk}\nabla_i\nabla_j\nabla_k +
 \frac{1}{2!} \nabla_i\, U_{2,2}^{ij}\nabla_j + U_{2,1}^i \nabla_i   \\
& := & U_{2,4} + U_{2,3} + U_{2,2} +U_{2,1} ,
\end{eqnarray}
and $U_{2,4}$ and $U_{2,2}$ are hermitian, while $U_{2,3}$ and $U_{2,1}$ are not.
In terms of the original $V_4$ and $V_3$, we have
\begin{eqnarray}
U_{2,4}^{ijkl} := V_4^{ijkl}, \
U_{2,3}^{ijk} := V_3^{ijk} -\frac{1}{2} \nabla_l V_4^{ijkl}, \
U_{2,2}^{ij} := - \frac{2}{4!} (\nabla_k\nabla_l V_4^{ijkl} ), \
U_{2,1}^i := \frac{1}{4!} (\nabla_j\nabla_k\nabla_l V_4^{ijkl} ). ~~~~~
\end{eqnarray}

\subsubsection{General case}
The change of the wave function $\psi = R^{(2)}\phi$ at $n=2$ is given by
 \begin{eqnarray}
R^{(2)} &=& R_1 ( 1 + R_2 ), \quad R_2 := R_{2,0} + R_{2,2},
\end{eqnarray}
where the $n=1$ term $R_1$ is already determined in the previous subsection, while the $n=2$ term $R_2$ contains the local function $R_{2,0}$ without derivatives and  $R_{2,2}$ with second derivatives as
\begin{eqnarray}
R_{2,2} &:=& \frac{1}{2!} R_{2,2}^{ij}\nabla_i\nabla_j .
\end{eqnarray}
As will be seen later, we can make $H^{(2)}$ hermitian without the first derivative term, $R_{2,1}$.   

The transformed Hamiltonian $\tilde H_2$ is given by
\begin{eqnarray}
\tilde H^{(2)} &:=& (R^{(2)})^{-1} H^{(2)} R^{(2)} \simeq (1-R_2) R_1^{-1} (H^{(1)} + U_2)  R_1(1+ R_2)\nonumber \\
&\simeq& \tilde H^{(1)}  + R_1^{-1} U_2 R_1 + \left[ \tilde H^{(1)}, R_2 \right], \quad
\tilde H^{(1)}= H_0 +\tilde V_0 + \tilde V_2 := \tilde V_0 + H_{1,2},
\label{eq:tilde_H1}
\end{eqnarray}
where $\tilde H^{(1)}$ is already made  hermitian by $R_1$ and
we neglect higher order terms such as $O\left(R_2^2\, \right)$ and $ O\left(R_2 H_2\, \right)$.

We first consider $\tilde U_2 := R_1^{-1} U_2 R_1$, which is evaluated as
\begin{eqnarray}
\tilde U_2 &=& U_{2,4} +\tilde U_{2,3} +\tilde U_{2,2} +\tilde U_{2.1} +\tilde U_{2,0},
\end{eqnarray}
where $\tilde U_{2,n}$ consists of $n$-th derivative terms, and $\tilde U_{2,2}$ can be taken to be hermitian while $\tilde U_{2,0}$ is always hermitian. (Note that $U_{2,4}$ is defined to be  hermitian.)  Explicit forms of $\tilde U_{2,n}$ in terms of $U_{2,l}$ are too complicated but unnecessary for our argument hereafter.

Similarly we can write
\begin{eqnarray}
X_0 := \left[ \tilde H^{(1)}, R_{2,0}\right] = X_{0,1} + X_{0,0}, \\
X_2 := \left[ \tilde H^{(1)}, R_{2,2}\right] = X_{2,3} +X_{2,2} + X_{2.,1} + X_{2,0},
\end{eqnarray}
where $n$ in $X_{k,n}$ represents the number of derivatives, and  $X_{k,2n}$  is taken to be hermitian.
Explicitly, we have
\begin{eqnarray}
X_{2,3} &:=& \frac{1}{3!} X_{2,3}^{ijk} \nabla_i\nabla_j\nabla_k, \quad
X_{2,2} := \frac{1}{2!}  \nabla_i X_{2,2}^{ij} \nabla_j, \quad
X_{2,1} := X_{2,1}^i \nabla_i, \\
X_{2,3}^{ijk} &=&  \left[ H_{1,2}^{il} (\nabla_l R_{2,2}^{jk}) -R_{2,2}^{il}(\nabla_l H_{1,2}^{jk})\right] + \mbox{2 permutations}, \\
X_{2,2}^{ij} &=&  - R_{2,2}^{il}(\nabla_k\nabla_l H_{1,2}^{kj}) 
 +\frac{1}{2}\left\{H_{1,2}^{kl}(\nabla_k\nabla_l R_{2,2}^{ij}) 
+(\nabla_k H_{1,2}^{kl})(\nabla_l R^{ij}_{2,2}) - R^{kl}_{2,2} (\nabla_k\nabla_l H_{1,2}^{ij}) \right\}, ~~~~~\\
X_{2,1}^i &=&- R_{2,2}^{ij}(\nabla_j \tilde V_0)
-\frac{1}{4}R_{2,2}^{kl}(\nabla_j\nabla_k\nabla_l H_{1,2}^{ij})-\frac{1}{2}(\nabla_j X_{2,2}^{ij}), 
\label{eq:X21}\\
X_{2,0}&=& -\frac{1}{2} R_{2,2}^{ij}(\nabla_i\nabla_j \tilde V_0),
\end{eqnarray}
and
\begin{eqnarray}
X_{0,1} &=& H_{1,2}^{ij} (\nabla_j R_{2,0} )\nabla_i , \qquad
X_{0,0}  = \frac{1}{2} H_{1,2}^{ij} (\nabla_i \nabla_j R_{2,0} ) 
+ \frac{1}{2} (\nabla_i H_{1,2}^{ij}) (\nabla_j R_{2,0} ) ,
\end{eqnarray}
where $H_{1,2}$ is defined in eq.~(\ref{eq:tilde_H1}).

The transformed Hamiltonian becomes
\begin{eqnarray}
\tilde H^{(2)} &=& \tilde H^{(1)} + U_{2,4} +\tilde U_{2,3} + \tilde U_{2,2} +\tilde U_{2,1} +\tilde U_{2,0} 
+ X_{2,3} + X_{2,2} + X_{2,1} + X_{2,0} + X_{0,1} + X_{0,0} .~~~~~~
\end{eqnarray}
To remove non-hermitian 3rd derivative terms, $R_{2,2}$ must satisfy 
\begin{eqnarray}
\tilde U_{2,3}^{ijk} + X_{2,3}^{ijk} &=& 0, \label{eq:3rd}
\end{eqnarray}
which becomes a linear 1st order partial differential equation for $R_{2,2}$.
Once $R_{2,2}$ is determined from this equation, $X_{2,2}$, $X_{2,1}$ and $X_{2,0}$ are completely fixed. To remove non-hermitian 1st derivative terms, $R_{2,0}$ must satisfy 
\begin{eqnarray}
\tilde U_{2,1}^i + X_{2,1}^i +  H_{1,2}^{ij} (\nabla_j R_{2,0} ) &=& 0,
\label{eq:1st}
\end{eqnarray}
which again becomes a linear 1st order partial differential equation for $R_{2,0}$, so that we can easily solve it to fix $X_{0,0}$.

We finally obtain
\begin{eqnarray}
\tilde H^{(2)} &=& \tilde H^{(2)} + U_{2,4} + \tilde U_{2,2} +\tilde U_{2,0}  + X_{2,2}  + X_{2,0}  + X_{0,0} ,
\end{eqnarray}
which is manifestly hermitian at $n=2$.

\subsubsection{$R_{2.2}$ and $R_{2.0}$ for the rotationally symmetric case}
In order to demonstrate that equations for $R_{2,2}$ and $R_{2,0}$ can be solved, we explicitly determine $R_{2,2}$ and $R_{2,0}$ for the rotationally symmetric case.

For this case, we have
\begin{eqnarray}
\tilde U_{2,3}^{ijk} &:=& V_{3a}(r) \hat r^i \hat r^j \hat r^k +  V_{3b}(r)\left\{ \hat r^i\delta^{jk} +
\hat r^j\delta^{ki} + \hat r^k\delta^{ij}\right\}, \quad \tilde U_{2,1}^i := V_1(r) \hat r^i, \\ 
H_{1,2}^{ij} &:=& H_{2a}(r) \hat r^i \hat r^j + H_{2b}(r) \delta^{ij}, \quad
R_{2,2}^{ij} := R_{2a}(r) \hat r^i \hat r^j + R_{2b}(r) \delta^{ij}, \quad R_{2,0} := R_{2,0}(r), 
\end{eqnarray}
which lead to
\begin{eqnarray}
X_{2,3}^{ijk} &=& 3 X_{3a}(r) \hat r^i \hat r^j \hat r^k +  X_{3b}(r)\left\{ \hat r^i\delta^{jk} +
\hat r^j\delta^{ki} + \hat r^k\delta^{ij}\right\}, \\
X_{3a} &:=&  (H_{2a} + H_{2b}) R_{2a}^\prime -  (R_{2a} + R_{2b}) H_{2a}^\prime 
-\frac{2}{r}\left(R_{2a} H_{2b} -H_{2a} R_{2b}\right), \\
X_{3b} &:=&  (H_{2a} + H_{2b}) R_{2b}^\prime -  (R_{2a} + R_{2b}) H_{2b}^\prime 
+ \frac{2}{r}\left(R_{2a} H_{2b} -H_{2a} R_{2b}\right). 
\end{eqnarray}
Thus eq.~(\ref{eq:3rd}) gives
\begin{eqnarray}
H_{2+} R_{2a}^\prime - H_{2a}^\prime R_{2+} -\frac{2}{r}\left(R_{2a} H_{2b} -H_{2a} R_{2b}\right) &=& \frac{V_{3a}}{3}, \\
H_{2+}  R_{2b}^\prime - H_{2b}^\prime R_{2+} +\frac{2}{r}\left(R_{2a} H_{2b} -H_{2a} R_{2b}\right)  &=& V_{3b}
\end{eqnarray}
with $R_{2+}:= R_{2a}+R_{2b}$ and $H_{2+}:=H_{2a}+H_{2b}$, which is simplified as
\begin{eqnarray}
H_{2+}  R_{2+}^\prime -H_{2+}^\prime  R_{2+}
 = \frac{ V_{3+}}{3}, \quad V_{3+} := V_{3a} + 3V_{3b} .
\end{eqnarray}
This equation can be easily solved as
\begin{eqnarray}
R_{2+}&=& C(r) H_{2+}(r),\qquad
C(r) := \int_{r_\infty}^r ds\, \frac{ V_{3+}(s)}{3 H_{2+}(s)^2}, 
\end{eqnarray}
where we assume the $s$-integral to be finite. In other words, singularities of the integrand between $ 0 < s < r_\infty$ are all integrable. 
From the original equations, individual terms are given by
\begin{eqnarray}
R_{2a}(r) &=& C(r) H_{2a}(r) + \frac{r^2}{3}\int_{r_\infty}^r ds\,  \frac{H_{2b}(s) V_{3a}(s) -3H_{2a}(s) V_{3b}(s)}{s^2 H_{2+}^2(s)},~~~~\\
R_{2b}(r) &=& C(r) H_{2b}(r) + \frac{r^2}{3}\int_{r_\infty}^r ds\,  \frac{3 H_{2a}(s)  V_{3b}(s) -H_{2b}(s) V_{3a}(s)}{s^2 H_{2+}^2(s)}.~~~~
\end{eqnarray}

Once $R_{2,2}(\vec r)$ is determined, eq.~(\ref{eq:X21}) fixes $X_{2,1}^i$ and eq.~(\ref{eq:1st}) becomes
\begin{eqnarray}
V_1(r) + X_1(r) + H_{2+}(r) R_{2,0}^\prime(r) &=& 0,  
\end{eqnarray}
which can be solved as
\begin{eqnarray}
R_{2,0}(r) &=&- \int_{r_\infty}^r ds\, \frac{V_1(s) +  X_1(s)}{ H_{2+}(s)}, 
\end{eqnarray}
where
\begin{eqnarray}
X_{2,1}^i &:=& X_1(r) \hat r^i,
\end{eqnarray}
and $X_1(r)$ is expressed in terms of $\tilde V_0$, $H_{2a}$, $H_{2b}$, $V_{3a}$ and $V_{3b}$.

\subsection{All orders}
We now argue that we can make the total Hamiltonian hermitian order by order in the derivative expansion.
The total Hamiltonian is given in the derivative expansion as
\begin{eqnarray}
H &=& H_0 + V_0 + \sum_{l=1}^\infty U_l, \qquad U_l := V_{2l} + V_{2l-1},
\end{eqnarray}
while the $n$-th order Hamiltonian is denoted as
\begin{eqnarray}
H^{(n)} &=& H_0 +V_0 +\sum_{l=1}^n U_l .
\end{eqnarray}
In previous subsections, we have already shown that $H^{(1)}$ and $H^{(2)}$ can be made hermitian.
As before, we make the even-derivative terms hermitian by introducing lower derivative terms, so that
\begin{eqnarray}
U_n &=& \sum_{k=1}^{2n} U_{n,k}, \qquad U_{n,2k} \mbox{ : hermitian},  
\end{eqnarray}
where $k$ of $U_{n,k}$ represents the number of derivatives in this terms, while $n$ corresponds to the order of this term. Throughout this subsection, we use the similar notations also for other quantities such as $F_{n,k}$, which is the $n$-th order term with $k$ derivatives, and is hermitian for even $k$.  

The transformation operator $R$ is expanded as
\begin{eqnarray}
R &=& R_1\left(1 + \sum_{l=2}^\infty R_l \right), 
\end{eqnarray}
where $R_n$ is expanded in terms of even numbers of derivatives as
\begin{eqnarray}
R_n &:=& \sum_{k=0}^{n-1} R_{n,2k} .
\end{eqnarray}

In order to prove that $H$ can be made hermitian order by order, we use the mathematical induction.
We have already seen that $H_1$ and $H_2$ can be made hermitian by $R_1$ and $R_1(1+R_2)$, respectively.

We next assume that $H_{n}$ can be  made hermitian by $R^{(n)} = R_1(1+\sum_{l=2}^n R_l)$ 
at the $n$-th order as
\begin{eqnarray}
\tilde H^{(n)} &:=& (R^{(n)})^{-1} H^{(n)} R^{(n)} \simeq \sum_{k=0}^{n} \tilde H^{(n)}_{2k} + \Delta \tilde H_{n+1},  
\label{eq:asume_n}
\end{eqnarray}
where $\tilde H^{(n)}_{2k}$ are all hermitian with $2k$ derivatives and contain terms whose orders are less than or equal to $n$, while $\Delta \tilde H_{n+1}$ is  non-hermitian at $(n+1)$-th order and consists of terms such as
\begin{equation*}
\left( \prod_{i=1}^s R_{k_i} \right)\times  R_1^{-1}U_l R_1
 \times \left(1 \mbox{ or } R_m \right), \quad  s=0,1,2,\cdots ,
\end{equation*}
for $1< k_i,l,m \le n$
with the constraint $\displaystyle \sum_{i=1}^s (k_i-1) + (l-1) +( m-1) =n$.
 Therefore, the maximum number of derivatives in $\Delta \tilde H^{(n+1)}$ is 
 $2 \displaystyle \sum_{i=1}^s (k_i-1) +2l + 2(m-1) =2(n+1)$, so that we can write
\begin{eqnarray}
\Delta \tilde H_{n+1} &=& \sum_{k=0}^{2(n+1)} \Delta\tilde H_{n+1,k}, 
\end{eqnarray}
where $k$ denotes the number of derivatives in $ \Delta\tilde H_{n+1,k}$, which is hermitian for even $k$.

We now consider the transformed Hamiltonian at the $(n+1)$-th order as
\begin{eqnarray}
 \tilde H^{(n+1)} &:=& \left(R^{(n+1)}\right)^{-1} H^{(n+1)}  R^{(n+1)} %\nonumber \\
 %&\simeq& 
 \simeq \left( R^{(n)}\right)^{-1} H^{(n)} R^{(n)} + \tilde U_{n+1} +\left[ \tilde H^{(1)}, R_{n+1}\right],
\end{eqnarray}
where the second term is evaluated as
\begin{eqnarray}
\tilde U_{n+1} &:=&R_1^{-1} U_{n+1} R_1 = \sum_{k=0}^{2(n+1)} \tilde U_{n+1,k},
\end{eqnarray}
while the 3rd term becomes 
\begin{eqnarray}
\left[ \tilde H^{(1)}, R_{n+1}\right] &=& \sum_{k=0}^{n} X_{n+1}[k], \quad
X_{n+1}[k] := \left[ \tilde H^{(1)}, R_{n+1,2k}\right] = \sum_{l=1}^{2k+1} X_{n+1,l}[k] .
\end{eqnarray}

Using the assumption of the mathematical induction in eq.~(\ref{eq:asume_n}), we have
\begin{eqnarray}
\tilde H^{(n+1)} &=& \sum_{k=0}^n \left( \tilde H^{(n)}_{2k} +\sum_{l=1}^k X_{n+1,2l}[k]\right)
+\sum_{k=0}^{n+1}\left( \Delta \tilde H_{n+1,2k} + \tilde U_{n+1,2k}\right) \nonumber \\
&+& \sum_{k=0}^n \left( \Delta \tilde H_{n+1,2k+1} +\tilde U_{n+1,2k+1} +\sum_{l=0}^k X_{n+1,2l+1}[k]\right).
\label{eq:Hn+1}
\end{eqnarray}
Using $n+1$ unknown $R_{n+1,2k}$ with $k=0,1,\cdots, n$, we can remove non-hermitian contributions in $\tilde H^{(n+1)}$ (the second line in eq.~(\ref{eq:Hn+1}) ), as shown below.

We first remove $(2n+1)$th order derivative terms in the second line by requiring 
\begin{eqnarray}
\Delta \tilde H_{n+1,2n+1} +\tilde U_{n+1,2n+1} + X_{n+1,2n+1}[n] &=& 0,
\end{eqnarray}
which fixes $R_{n+1,2n}$.

The next condition becomes
\begin{eqnarray}
\Delta \tilde H_{n+1,2n-1} +\tilde U_{n+1,2n-1} + X_{n+1,2n-1}[n] + X_{n+1,2n-1}[n-1] &=& 0,
\end{eqnarray}
where $X_{n+1,2n-1}[n] $ is already  determined completely from $R_{n+1,2n}$. 
Therefore, the above equation determines $R_{n+1,2n-2}$ in $X_{n+1,2n-1}[n-1]$. 

Repeating this procedure, we can remove all non-hermitian contributions as follows.
For general $k=0,1,2,\cdots, n$, we have
\begin{eqnarray}
\Delta \tilde H_{n+1,2k+1} +\tilde U_{n+1,2k+1} + \sum_{l=k+1}^n X_{n+1,2k+1}[l] + X_{n+1,2k-1}[k] &=& 0,
\end{eqnarray}
where $ \displaystyle \sum_{l=k+1}^n X_{n+1,2k+1}[l] $ has already been determined from $R_{n+1,2l}$ with $l=n,n-1, n-2,\cdots, k+1$. Thus the above condition fixes $R_{n+1,2k}$ in $X_{n+1,2k-1}[k]$.
Therefore it is shown that
$H^{(n+1)}$ can be made hermitian as
\begin{eqnarray}
\tilde H^{(n+1)} &=&  \sum_{k=0}^n \left( \tilde H^{(n)}_{2k} +\sum_{l=1}^k X_{n+1,2l}[k]\right)
+\sum_{k=0}^{n+1}\left( \Delta \tilde H_{n+1,2k} + \tilde U_{n+1,2k}\right) .
\end{eqnarray}

The proof that non-hermitian $H$ can be made hermitian order by order
is thus completed by the mathematical induction. 

We note here that the n-th order Hamiltonian, given by Eqs.~(\ref{eq:V_n}) and (\ref{eq:U_n}), contains 
$2n+1$ new  unknown functions to be extracted from the NBS wave functions generated by lattice QCD calculations.
Thus, in order to perform the $n$-th order analysis, we totally need $(n+1)^2$ NBS wave functions, which must be independent beyond numerical uncertainties.   
This may give rise to severe limitations in applying the present formalism to the cases where the higher order terms are important. In the next section we will see that, in the case of $\Xi \Xi (^{1}S_{0})$ scattering, the LO potential already gives a good approximation for the scattering phase shift, while the NLO corrections to the phase shift gradually appear as the energy increases.

\section{Hermitizing the NLO potential for the $\Xi\Xi(^1S_0)$ system}
\label{sec:application}
In this section,  
we actually apply the method in Sec.~\ref{sec:hermitizing}
to lattice QCD data.
We consider the $\Xi\Xi(^1S_0)$ system, whose potential is much more precise than those
for $NN$ or $N\Lambda$. thanks to more strange quarks in the system,
in order to make the NLO analysis numerically possible.
The potential for the $\Xi\Xi(^1S_0)$ was calculated on 2+1 flavor QCD ensembles\cite{Yamazaki:2012hi} at $ a= 0.09$ fm on a $64^4$ lattice 
with heavy up/down quark masses and the physical strange quark mass, $m_\pi = 0.51$ GeV, $m_K = 0.62$ GeV, $m_N = 1.32$ GeV and $m_\Xi =1.46$ GeV.
See Ref.~\cite{Iritani:2018zbt} for more details.

The rotationally symmetric potential in the previous section for the $n=1$ case can be rewritten as
\begin{eqnarray}
V(\vec r, \nabla) &=& V_0(r) + V_1(r) \, \hat r^i \nabla_i +V_2(r) \nabla^2 + V_3(r) \vec L^2,
\end{eqnarray}
where the expression in the previous section is recovered by replacing  
\begin{eqnarray}
V_1(r) &\rightarrow V_1(r) -\dfrac{V_{2a}(r)}{r}, \quad
V_2 (r) \rightarrow \dfrac{1}{2}\left( V_{2a}(r) + V_{2b}(r) \right),\quad
V_3(r) \rightarrow \dfrac{V_{2a}(r)}{2r^2}.
\end{eqnarray}
Since the $V_{3} (r)$ term does not contribute to the S wave scattering, we ignor this term in the present analysis.
Having only two NBS wave functions, one from the wall source, the other from the smeared source, available from the previous calculation~\citep{Iritani:2018zbt},
we consider two different extractions of potentials, one with $V_1(r)=0$ (NLO$_{\rm A}$), the other with $V_2(r)=0$ (NLO$_{\rm B}$), in the present analysis.

\subsection{NLO$_{\rm A}$: NLO analysis without $V_1$ }
\begin{figure}
  \centering
 \includegraphics[width=0.49\textwidth,clip]{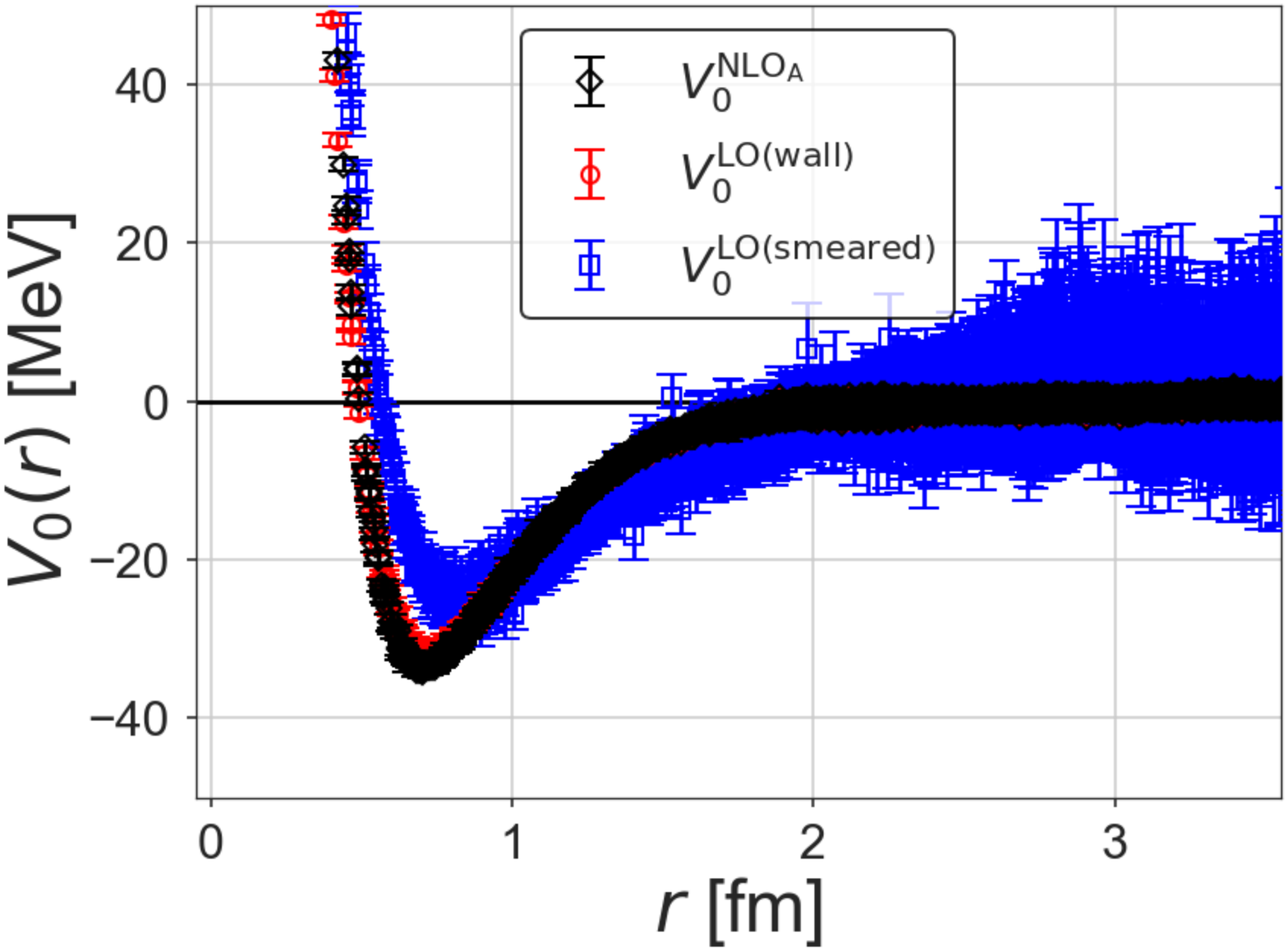}
 \includegraphics[width=0.49\textwidth,clip]{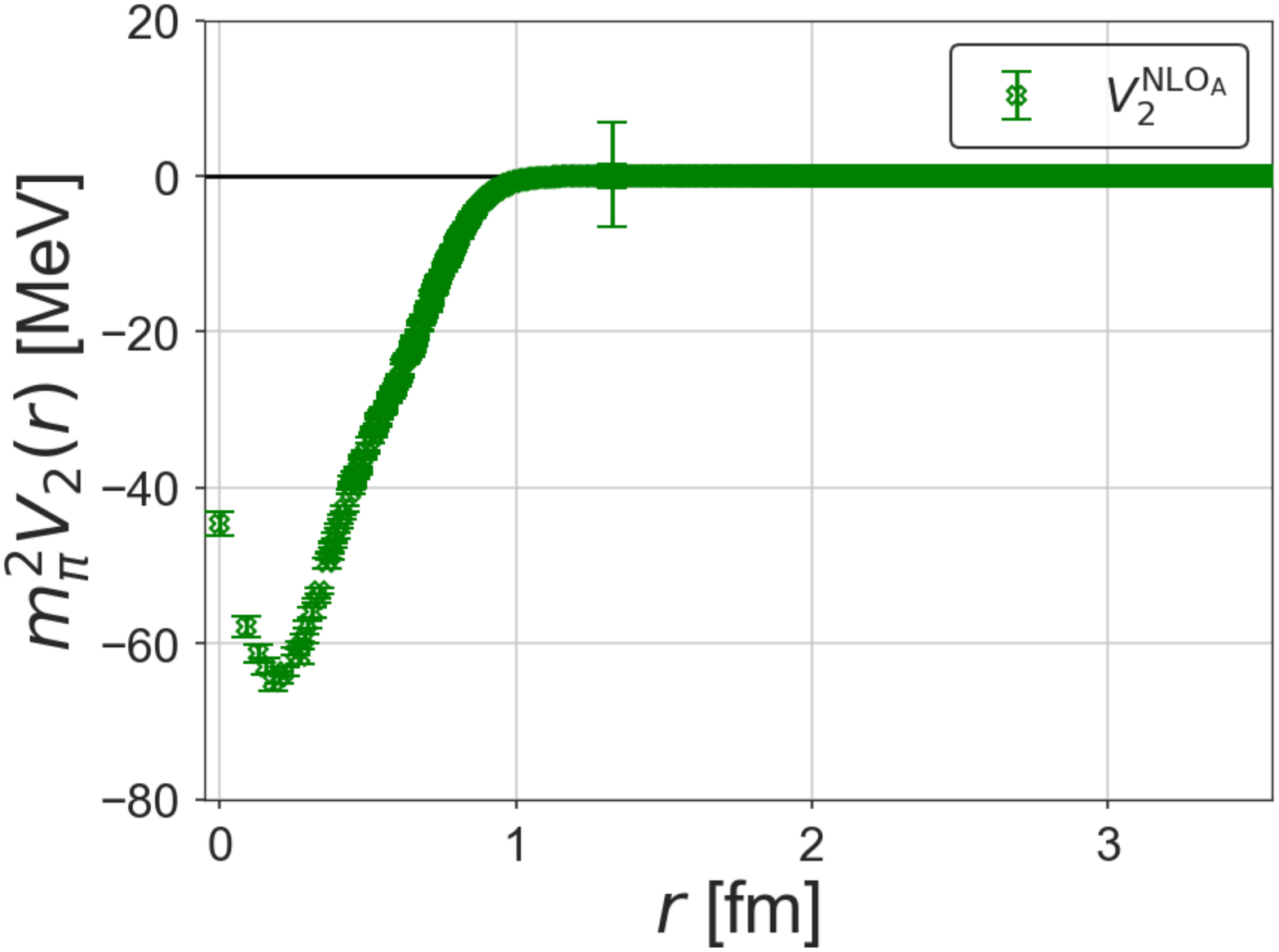}
  \caption{
    \label{fig:NLO_A}
       NLO potentials without $V_1$ (NLO$_{\rm A}$): (Left) $V^{\rm NLO_A}_0(r)$ (black open diamonds), together wth  $V^{\rm LO}_0(r)$ from the wall source (red open circles) and the smeared source (blue open squares). (Right) $V^{\rm NLO_A}_2(r)$ multiplied by $m_\pi^2$ (green open crosses) . 
 }
\end{figure}
We first consider the NLO$_{\rm A}$ potential for  $\Xi\Xi(^1S_0)$, 
\begin{eqnarray}
V^{\rm NLO_A}(\vec r, \nabla) &=& V^{\rm NLO_A}_0(r) + V^{\rm NLO_A}_2(r) \nabla^2, 
\end{eqnarray}
where $V^{\rm NLO_A}_0(r)$, together with
$V^{\rm LO(wall)}_0(r)$ from the wall source and $V^{\rm LO(smeared)}_0(r)$ from the smeared source, 
are plotted in Fig.~\ref{fig:NLO_A} (Left), and
$m_\pi^2 V^{\rm NLO_A}_2(r)$ is given in Fig.~\ref{fig:NLO_A} (Right).
Small differences between $V^{\rm NLO_A}_0(r)$ and $V^{\rm LO(wall)}_0(r)$ are observed
at short distance, due to contributions from  $V^{\rm NLO_A}_2(r)$, which is non-zero only at $r < 1$ fm.

\begin{figure}[tb]
  \centering
 \includegraphics[width=0.49\textwidth,clip]{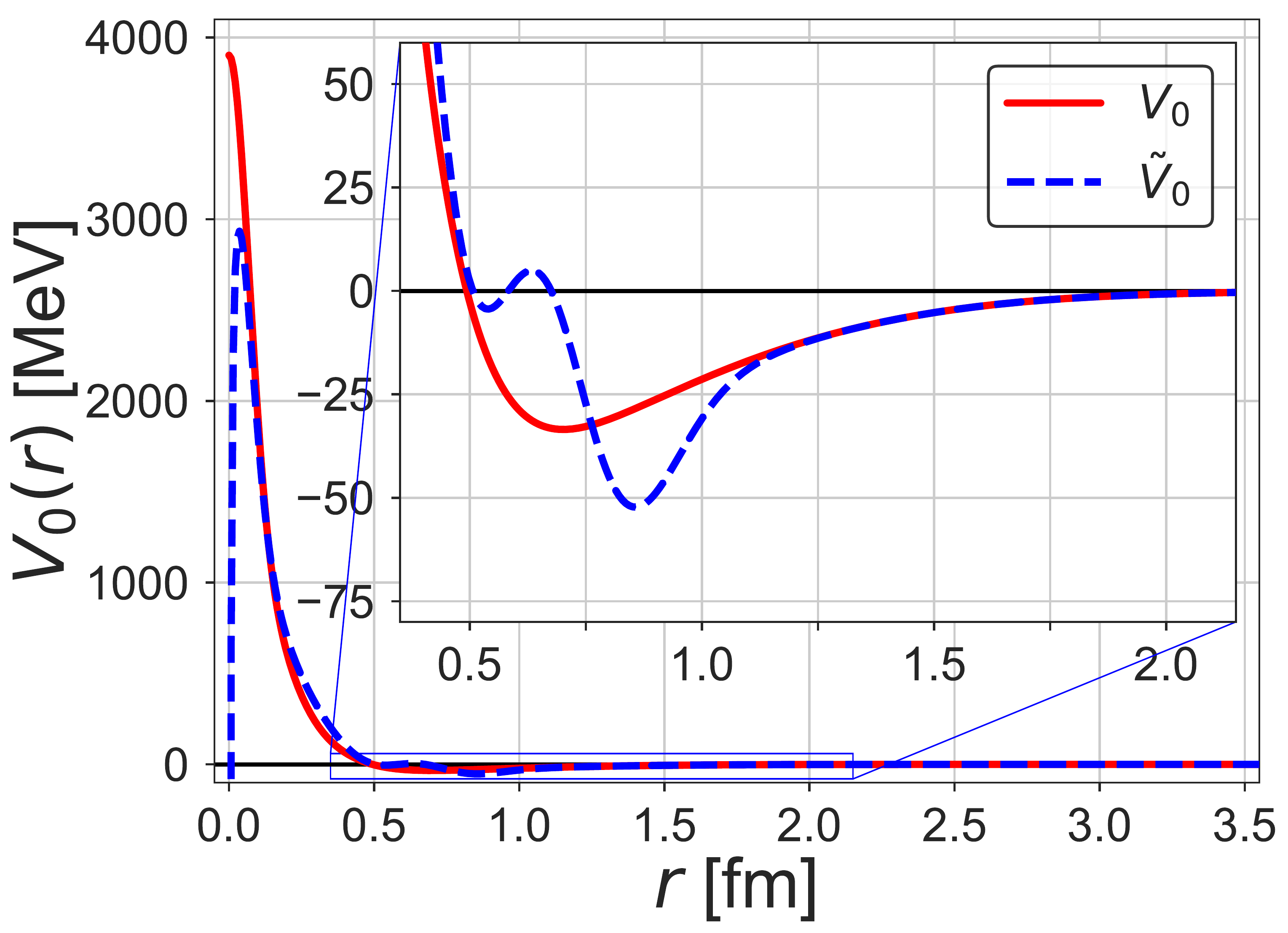}
 \includegraphics[width=0.49\textwidth,clip]{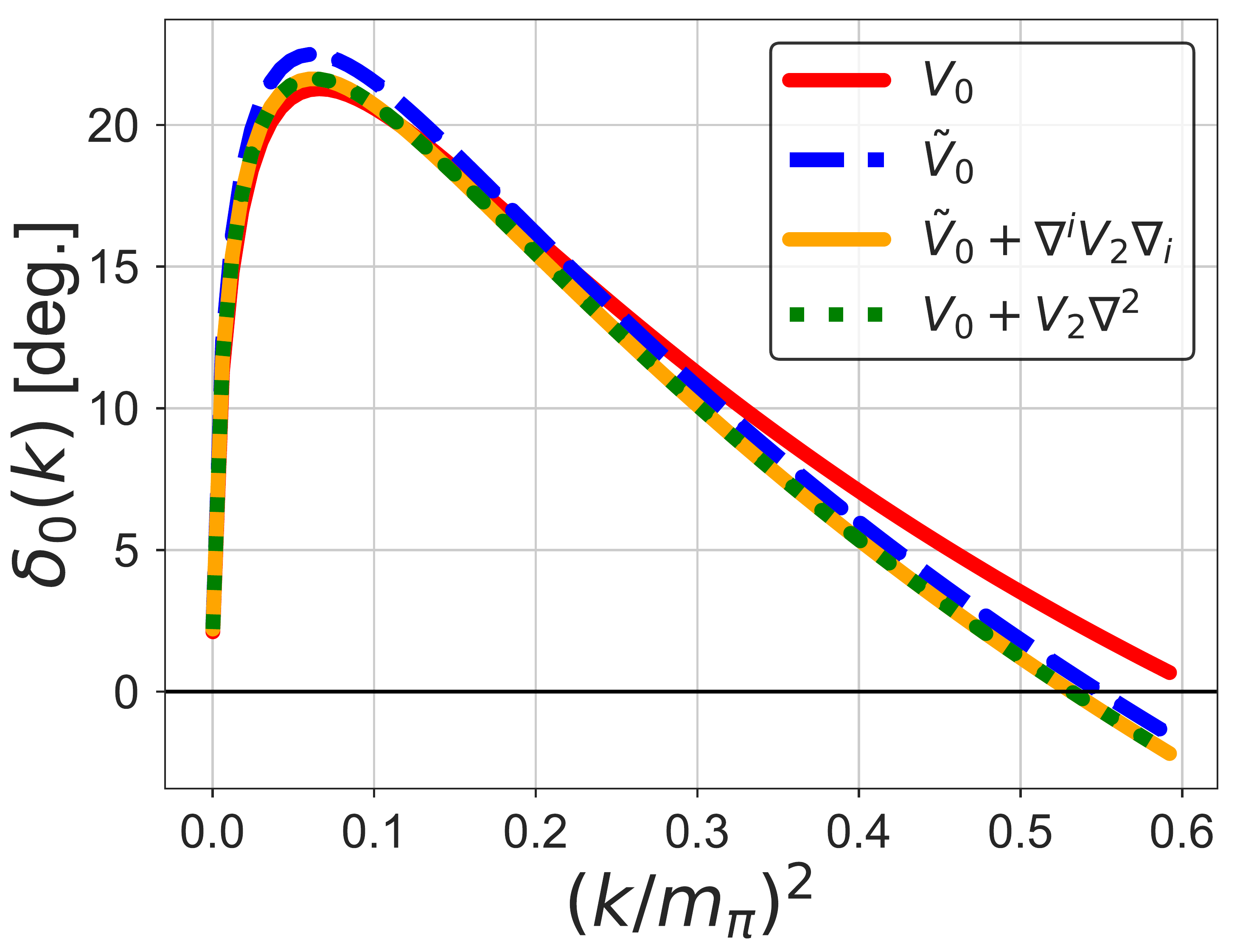}
  \caption{
    \label{fig:Hermitize}
      (Left) $\tilde V^{\rm NLO_A}_0(r)$ (blue dashed line) for $\Xi\Xi(^1S_0)$ together with $V^{\rm NLO_A}_0(r)$ (red solid line).  (Right) Scattering phase shifts for  $\Xi\Xi(^1S_0)$
       from $V^{\rm NLO_A}_0$ (red solid line), $\tilde V^{\rm NLO_A}_0$ (blue dashed line), $V^{\rm NLO_A}(\vec r,\nabla)$ (green dotted line), and $\tilde V^{\rm NLO_A}(\vec r,\nabla)$ (yellow solid line). 
        }
\end{figure}
According to the procedure in Sec.~\ref{sec:hermitizing}, we can make this non-hermitian potential hermitian as
\begin{eqnarray}
\tilde V^{\rm NLO_A}(\vec r, \nabla) &=& \tilde V_0^{\rm NLO_A}(r) + \nabla^i V^{\rm NLO_A}_2(r) \nabla_i,
\end{eqnarray}
where
\begin{eqnarray}
\tilde V^{\rm NLO_A}_0(r) &=& V^{\rm NLO_A}_0(r) +\frac{(V^{\rm NLO_A}_2)^\prime(r)}{r} + \frac{(V^{\rm NLO_A}_2)^{\prime\prime}(r)}{2} +\frac{m_\Xi}{4}\frac{ \{(V^{\rm NLO_A}_2)^\prime(r)\}^2}{1-m_\Xi V^{\rm NLO_A}_2(r)} .
\end{eqnarray}
{Here $\tilde V_0$ is calculated from the fitted functions for  $V^{\rm NLO_A}_0(r)$ and $V^{\rm NLO_A}_2(r)$ in Ref.~\cite{Iritani:2018zbt}, together with 
the 1st and second derivatives of $V_2$ analytically calculated from the fitted function. 
We compare $\tilde V^{\rm NLO_A}_0(r)$ with $V^{\rm NLO_A}_0(r)$ in 
 Fig.~\ref{fig:Hermitize} (Left), while
we plot the phase shifts of the $\Xi\Xi(^1S_0)$ obtained with $\tilde V^{\rm NLO_A}_0(r)$, $V^{\rm NLO_A}_0(r)$,  $V^{\rm NLO_A}(\vec r,\nabla)$  and $\tilde V^{\rm NLO_A}(\vec r,\nabla)$ in Fig.~\ref{fig:Hermitize} (Right).
For visibility, only central values are given here. 
Errors of $\tilde V_0^{\rm NLO_A}$ are comparable to those of $V_0^{\rm NLO_A}$,
which can be seen in Fig.~\ref{fig:NLO_A} (Left).
We notice that the leading order term $V^{\rm NLO_A}_0$ of the original non-hermitian NLO$_{\rm A}$ potential describes the behavior  of the scattering phase shift rather well at low energy but show some deviation from the NLO$_{\rm A}$ result at high energies, while the local term  $\tilde V^{\rm NLO_A}_0$ of the hermitized NLO$_{\rm A}$ potential describes the scattering phase shift in a wider energy ranges, $ 0 \le (k/m_\pi)^2 \le 0.6$ with small deviations around turning point at 
$(k/m_\pi)^2\simeq 0.1$. Of course, non-hermitian NLO$_{\rm A}$ potential $V^{\rm NLO_A}(\vec r,\nabla)$ and the hermitized NLO$_{\rm A}$ potential $\tilde V^{\rm NLO_A}(\vec r,\nabla)$
give identical results  at all energies by construction.

\subsection{NLO$_{\rm B}$: NLO analysis without $V_2$}
\begin{figure}
  \centering
 \includegraphics[width=0.49\textwidth,clip]{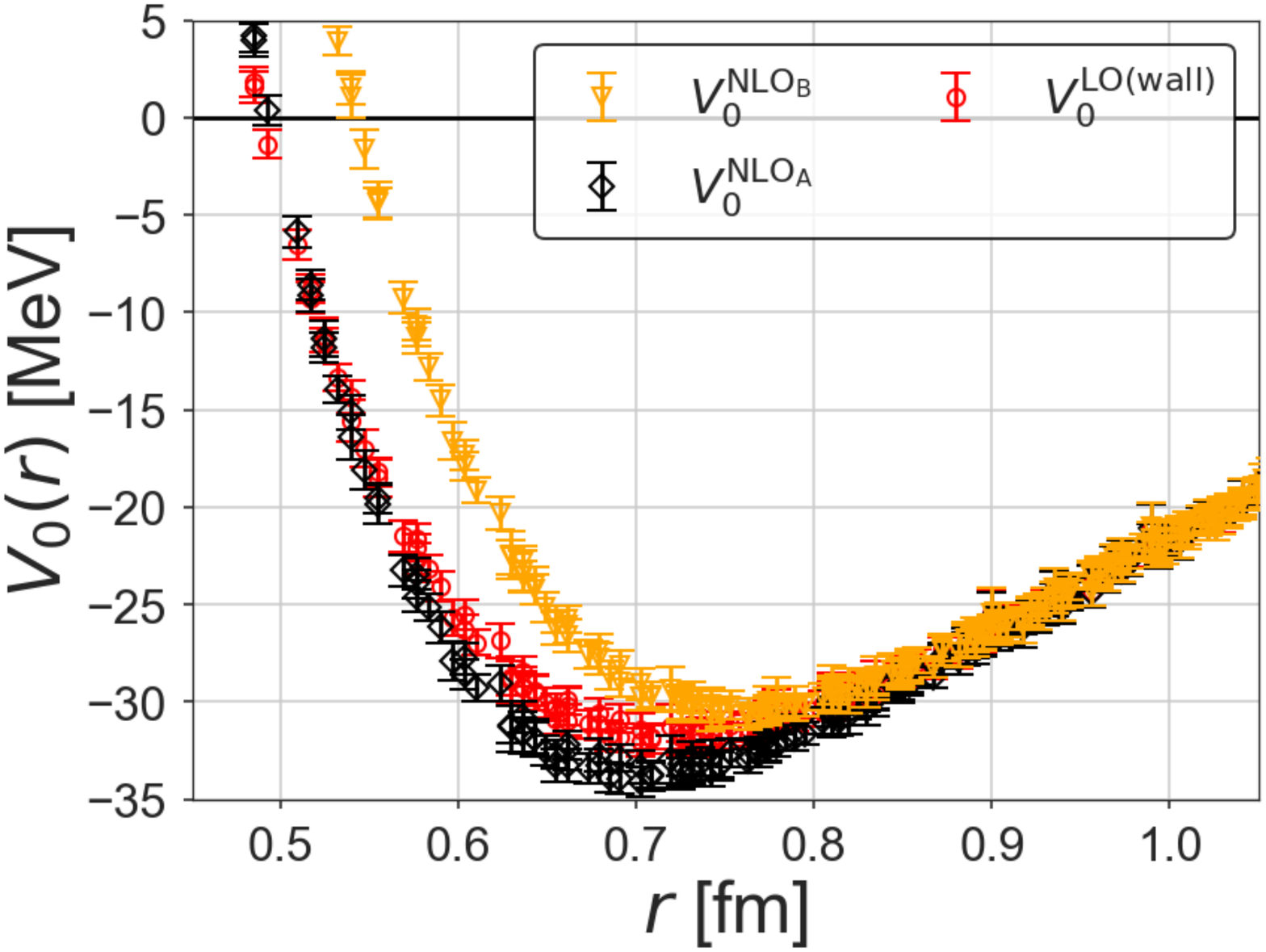}
 \includegraphics[width=0.49\textwidth,clip]{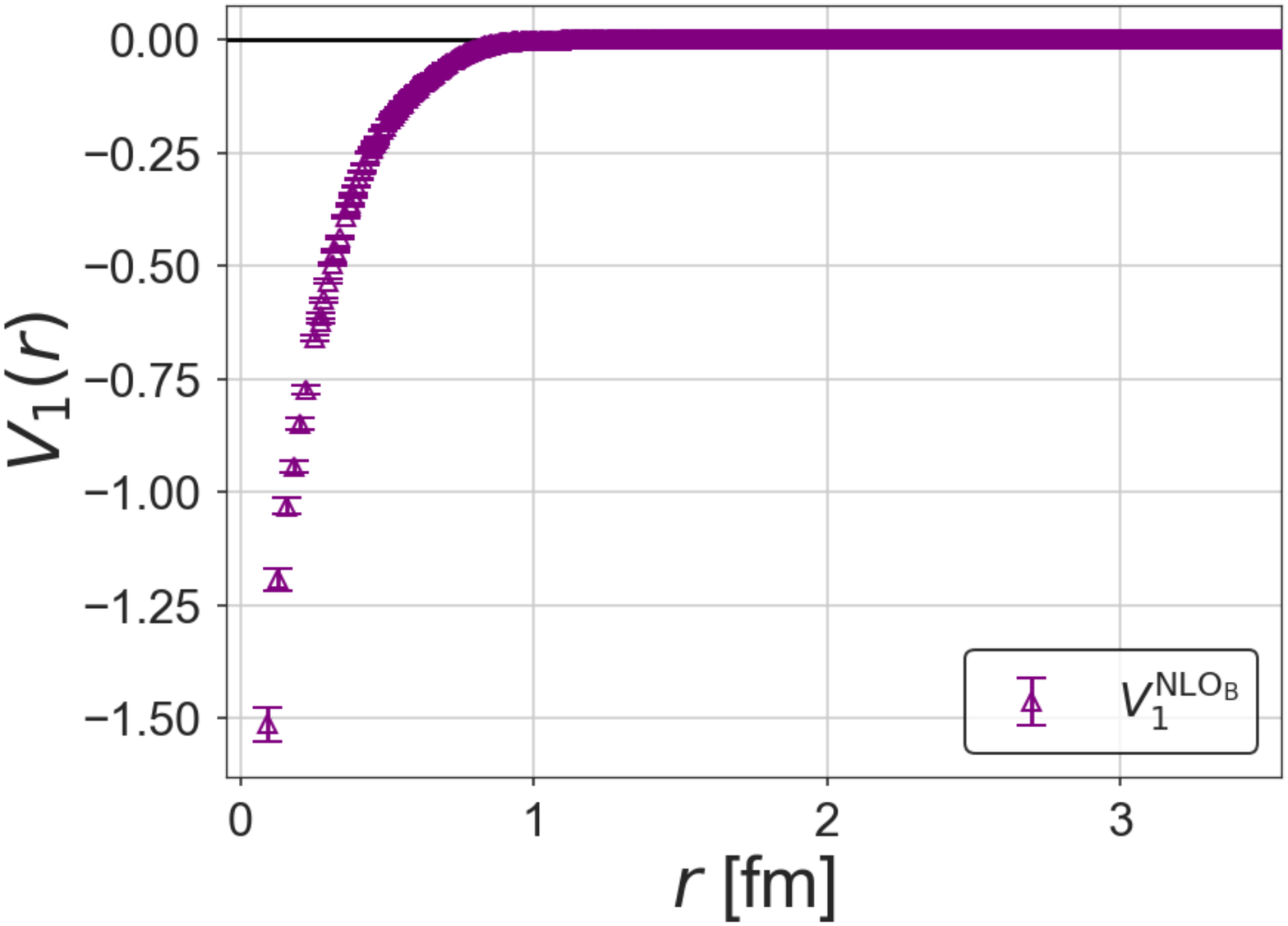}
  \caption{
    \label{fig:NLO_B}
      (Left) $V_0^{\rm NLO_B}(r)$ (yellow open triangles) as a function of $r$ at short distances,
      together with $V_0^{\rm LO (wall)}(r)$ (red open circles) and $V_0^{\rm NLO_A}(r)$ (black open triangles). (Right) $V_1^{\rm NLO_B}(r)$ (purple open triangles).         
  }
\end{figure}
We next consider the NLO analysis without $V_2$ (NLO$_{\rm B}$), whose potential is given by
\begin{eqnarray}
V^{\rm NLO_B}(\vec r, \nabla) &=& V^{\rm NLO_B}_0(r) + V^{\rm NLO_B}_1(r)\, \hat r^i \nabla_i .
\end{eqnarray}
Fig.~\ref{fig:NLO_B} (Left) shows $V_0^{\rm NLO_B}(r)$ (yellow open triangles), together with
$V_0^{\rm LO (wall)}(r)$ (red open circles) and $V_0^{\rm NLO_A}(r)$ (black open triangles),
at 0.45 fm $ \le r \le $1.05 fm, 
while Fig.~\ref{fig:NLO_B} (Right) gives  $V_1^{\rm NLO_B}(r)$ (purple open triangles)
at  $r \le 3.5$ fm.
Unlike $V_0^{\rm NLO_A}(r)$, $V_0^{\rm NLO_B}(r)$ deviates largely from $V_0^{\rm LO (wall)}(r)$ at short distances.

According to the procedure in Sec.~\ref{sec:hermitizing}, we convert this non-hermitian NLO$_{\rm B}$ potential to a local hermitian potential $\tilde V_0^{\rm NLO_B}(r)$, where
\begin{eqnarray}
\tilde V^{\rm NLO_B}_0(r) &=& V_0^{\rm NLO_B}(r) -\frac{ V_1^{\rm NLO_B}(r)}{r} -\frac{1}{2}  (V_1^{\rm NLO_B})^\prime (r) +  \frac{m_\Xi}{4} (V_1^{\rm NLO_B}(r) )^2 ,
\end{eqnarray} 
which is shown in Fig.~\ref{fig:delta_NLO_B} (Left) by blue dotted line, together with $V_0^{\rm LO (wall)}(r)$ (red solid line) and $V_0^{\rm NLO_B}(r)$ (yellow dashed line). 
The attractive pocket of $\tilde V_0^{\rm NLO_B}(r)$ around $r\simeq 0.6$ fm is much deeper than that of $V_0^{\rm LO (wall)}(r)$ or $V_0^{\rm NLO_B}(r)$.
Fig.~\ref{fig:delta_NLO_B} (Right) compares the scattering phase shifts $\delta_0(k)$ among
$V_0^{\rm LO (wall)}(r)$ (red solid circles),  $V_0^{\rm NLO_B}(r)$ (yellow solid down-triangles),
$V_0^{\rm NLO_A}(r)$ (black solid diamonds), $V^{\rm NLO_B}(\vec r, \nabla)$ (purple solid up-triangles), $V^{\rm NLO_A}(\vec r, \nabla)$ (green crosses) and $\tilde V_0^{\rm NLO_B}(r)$ (blue open squares).
By construction,  $\tilde V_0^{\rm NLO_B}(r)$ and $V^{\rm NLO_B}(\vec r, \nabla)$ give identical results,
which also agree well with $\delta_0(k)$ from  $V_0^{\rm NLO_A}(r)$.  
As mentioned before, $V_0^{\rm LO (wall)}(r)$ and $V_0^{\rm NLO_A}(r)$ give
good approximations at low energy but show small deviations as energies increase.
However $\delta_0(k)$ from $V_0^{\rm NLO_B}(r)$ deviates from others even at low energies.
Finally it is noted that two different NLO potentials, $V^{\rm NLO_A}(\vec r,\nabla)$ (green solid crosses) and  $V^{\rm NLO_B}(\vec r,\nabla)$ (purple solid up-triangles),  agree well even at high energies, though the first derivative term ($V_1^{\rm NLO_B}(r)$) has larger effects than the second derivative term ($V_2^{\rm NLO_A}(r)$)  on the scattering phase shift.
\begin{figure}
  \centering
 \includegraphics[width=0.49\textwidth,clip]{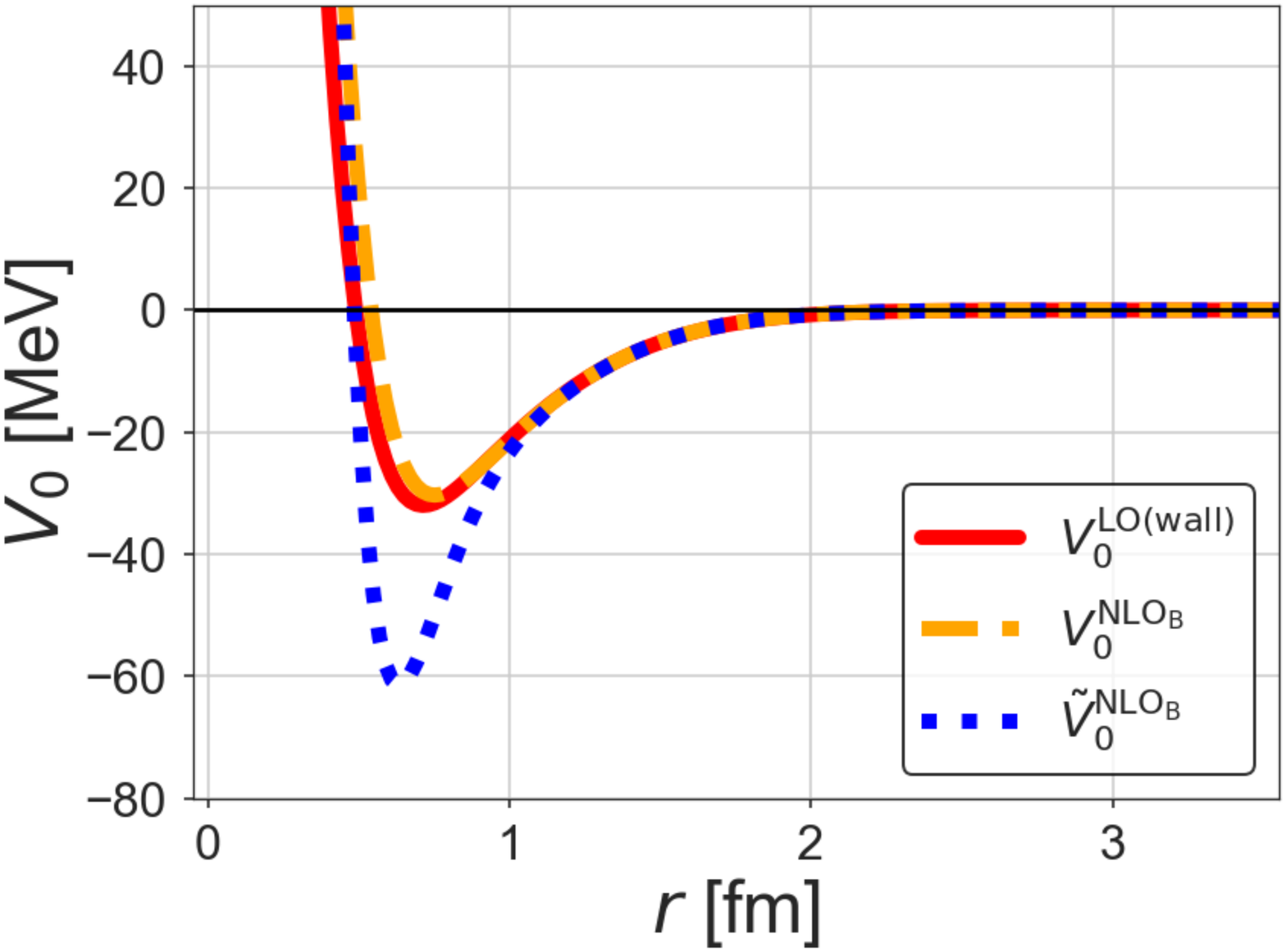}
 \includegraphics[width=0.49\textwidth,clip]{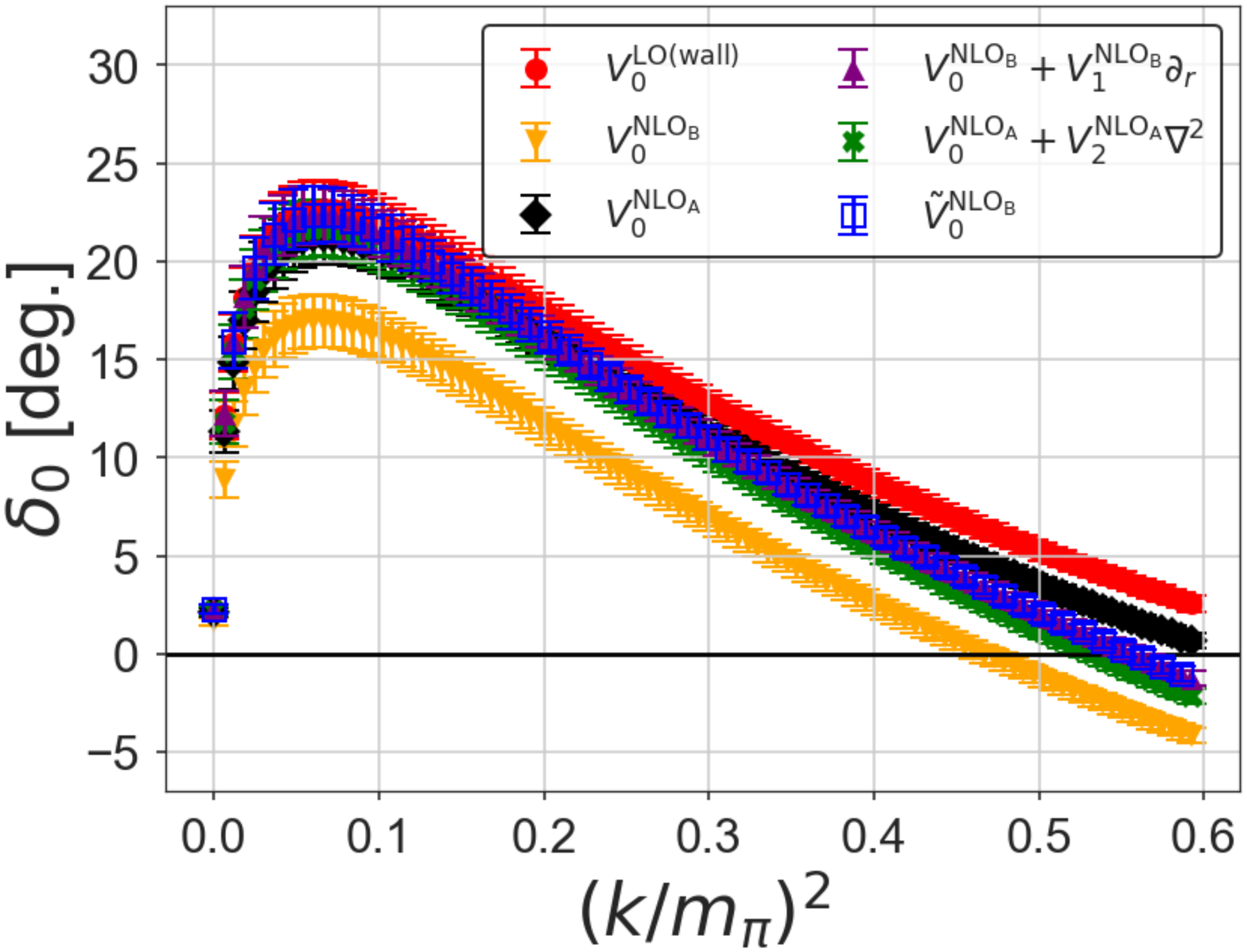}
  \caption{
    \label{fig:delta_NLO_B}
      (Left) $\tilde V_0^{\rm LO_B}(r)$ (blue dotted line) for $\Xi\Xi(^1S_0)$ together with $V_0^{\rm LO (wall)}(r)$ (red solid line) and $V_0^{\rm NLO_B}(r)$ (yellow dashed line).  
      (Only central values are plotted for visibility.)
      (Right) Scattering phase shifts $\delta_0(k)$ for  $\Xi\Xi(^1S_0)$
       from $V_0^{\rm LO (wall)} (r)$ (red solid circles), $V_0^{\rm NLO_B}(r)$ (yellow solid down-triangles), $V_0^{\rm NLO_A}(r)$ (black solid diamonds), $V^{\rm NLO_B}(\vec r, \nabla)$ (purple solid up-triangles), $V^{\rm NLO_A}(\vec r, \nabla)$ (green crosses) and $\tilde V_0^{\rm NLO_B}(r)$ (blue open squares).
       }
\end{figure}

\section{Summary and concluding remarks}
\label{sec:summary}
The HAL QCD potential expressed as an energy independent non-local potential is known to be non-hermitian due to the nature of the Nambu Bethe Salpeter (NBS) wave function used to extract it:
While the leading order (LO) term in the derivative expansion of the potential is local and hermitian, the higher order terms are in general non-hermitian.
In this paper, we have formulated a way of hermitizing it in the derivative expansion.  Since the hermitized potential can be expressed to contain only even number of derivatives, we classify the first and second order derivative terms as the next-to-leading order (NLO) and in general $(2n-1)$ and $2n$ derivative terms as the n-th order. Starting from the NLO terms, which can be made hermitian exactly, we have shown that the higher order terms can be hermitized order by order to all orders using the mathematical induction in the derivative expansion.

In order to see the feasibility of our formalism, we applied it to the case of $\Xi \Xi (^{1}S_{0})$ scattering for which two independent NBS wave functions were available from the lattice QCD calculations\cite{Iritani:2018zbt}. 
Since two NBS wave functions are insufficient for the full NLO analysis which requires three unknown functions, $V_{0}(r)$, $V_{1}(r)$ and $V_{2}(r)$,
we carried out two NLO analyses, one without $V_{1}$ $(\rm{NLO_{A}})$ and the other without $V_{2}$ $(\rm{NLO_{B}})$. Although the two hermitized potentials, $\bar{V}^{\rm{NLO_{A}}}$ and $\bar{V}^{\rm{NLO_{B}}}$, look very different, the former containing a second order derivative term while the latter being purely local, they give essentially the same phase shifts within the uncertainties of the calculations. This agreement indicates that the obtained NLO phase shift can be regarded approximately as the yet unknown exact one. 
By comparing it to the LO phase shifts obtained in ref.~\cite{Iritani:2018zbt}, we find that the LO phase shift from the NBS wave function with the wall source is very similar to the NLO phase shift at low energies while it is slightly larger at higher energies.  The LO analysis with the wall source is thus well justified for  the $\Xi \Xi (^{1}S_{0})$ scattering. 

While the non-hermitian potential is fine  as long as we are interested in the two-body observables such as scattering amplitudes and binding energies, 
the hermitian version is more convenient for a comparison  with phenomenological interactions and also for using it as a two-body interaction in many-body systems.  

\appendix

\section{Convergence of the derivative expansion for non-local potentials}
\label{app:conv}
In this appendix, we briefly discuss an issue on the convergence of the derivative expansion
for non-local potentials. 

Let us consider a non-local potential $V({\bf x},{\bf y})$, which can be expressed in terms of the derivative expansion as
\begin{equation}
V({\bf x},{\bf y}) = \sum_{n=0}^\infty \frac{1}{n!} V_l^{i_1i_2\cdots i_n}({\bf x})
\nabla_{i_1}\nabla_{i_2}\cdots \nabla_{i_n} \delta^{(3)}({\bf x} - {\bf y}),
\label{eq:derivative}
\end{equation}
where no symmetry is assumed for generality.

Applying this potential to a plane wave $e^{i {\bf k}\cdot {\bf x}}$, we obtain
\begin{eqnarray}
\bar V({\bf x}, {\bf k}) e^{i {\bf k}\cdot {\bf x}} &\equiv & \int d{\bf y}\, V({\bf x},{\bf y}) e^{i {\bf k}\cdot {\bf y}} = \sum_{n=0}^\infty \frac{i^n}{n!} V_l^{i_1i_2\cdots i_n}({\bf x}) k_{i_1} k_{i_2} \cdots k_{i_n}
e^{i {\bf k}\cdot {\bf x}} 
\label{eq:expand1}\\
&=& e^{i {\bf k}\cdot {\bf x}} \int d{\bf r} \,V({\bf x},{\bf x}+{\bf r}) e^{i {\bf k}\cdot {\bf r}}
= e^{i {\bf k}\cdot {\bf x}} \sum_{n=0}^\infty \int d{\bf r}\, V({\bf x},{\bf x}+{\bf r})\frac{i^n}{n!}
({\bf k}\cdot {\bf r})^n .
\label{eq:expand2}
\end{eqnarray}
By equating (\ref{eq:expand1}) and (\ref{eq:expand2}), we have
\begin{eqnarray}
V_l^{i_1i_2\cdots i_n}({\bf x}) &=& \int d{\bf r} V({\bf x},{\bf x}+{\bf r}) r_{i_1} r_{i_2}\cdots r_{i_n}
\equiv  V_{l}({\bf x}) \langle \, r_{i_1} r_{i_2} \cdots r_{i_n}\rangle_{\bf x} ,
\end{eqnarray}
where we define the $n$-th moment of non-locality, $ \langle r_{i_1} r_{i_2} \cdots r_{i_n}\rangle_{\bf x}$,  satisfying
$ \langle 1 \rangle_{\bf x} =1$, and $V_l({\bf x})$ is the term at $n=0$ (the local term)
in the derivative expansion eq.~(\ref{eq:derivative}). 
Then $\bar V({\bf x}, {\bf k})$ can be expanded as
\begin{eqnarray}
\bar V({\bf x}, {\bf k} ) &=& V_l({\bf x} )\, \sum_{n=0}^\infty \frac{i^n}{n!} \, \langle ({\bf k}\cdot{\bf r})^n\rangle_{\bf x} .
\label{eq:expand3}
\end{eqnarray}

The convergence of the sum in eq.~(\ref{eq:expand3}) is guaranteed if the absolute magnitude of the $n$-th moment of non-locality grows slower than $n^{\alpha n}$ with $\alpha < 1$ as $n\rightarrow\infty$, though the convergence rate depends on  ${\bf x}$ and ${\bf k}$,
reflecting the detail for the non-locality of $V({\bf x},{\bf x} +{\bf r})$.
It is clear that the sum converges as ${\bf k}\rightarrow 0$.
The convergence of the derivative expansion applied to a wave function, which can be expressed as a superposition of plane waves, is usually guaranteed, 
though the convergence rate depends on the nature of the wave function. 
In practice, 
for example, the importance of the NLO terms can be estimated from
$\vert {\bf k} \vert \vert \langle {\bf r}\rangle_{\bf x}\vert$ or $ {\bf k}^2 \vert \langle {\bf r}^2\rangle_{\bf x}\vert$,  where ${\bf k}$ is  the local wave number.

If the total energy is small enough,
the wave function is expanded by plane waves with small $\vert {\bf k}\vert $ only.
As the total energy increases, higher order terms become important to cause slower convergence,
and even may cease to converge at some energy,
depending on the non-locality of  $V({\bf x},{\bf x} +{\bf r})$.
In this paper, we assume that the non-locality of potentials is mild enough for the derivative expansion to converge.

\section*{Acknowledgements}
S. A. is supported in part by the Grant-in-Aid of the Japanese Ministry of Education, Sciences and Technology, Sports and Culture (MEXT) for Scientific Research (Nos. JP16H03978, JP18H05236),  
by a priority issue (Elucidation of the fundamental laws and evolution of the universe) to be tackled by using Post ``K" Computer, and by Joint Institute for Computational Fundamental Science (JICFuS). 
 
\bibliography{HALQCD}

\begin{thebibliography}{37}
\expandafter\ifx\csname natexlab\endcsname\relax\def\natexlab#1{#1}\fi
\expandafter\ifx\csname bibnamefont\endcsname\relax
  \def\bibnamefont#1{#1}\fi
\expandafter\ifx\csname bibfnamefont\endcsname\relax
  \def\bibfnamefont#1{#1}\fi
\expandafter\ifx\csname citenamefont\endcsname\relax
  \def\citenamefont#1{#1}\fi
\expandafter\ifx\csname url\endcsname\relax
  \def\url#1{\texttt{#1}}\fi
\expandafter\ifx\csname urlprefix\endcsname\relax\def\urlprefix{URL }\fi
\providecommand{\bibinfo}[2]{#2}
\providecommand{\eprint}[2][]{\url{#2}}

\bibitem[{\citenamefont{Luscher}(1991)}]{Luscher:1990ux}
\bibinfo{author}{\bibfnamefont{M.}~\bibnamefont{Luscher}},
  \bibinfo{journal}{Nucl. Phys.} \textbf{\bibinfo{volume}{B354}},
  \bibinfo{pages}{531} (\bibinfo{year}{1991}).

\bibitem[{\citenamefont{Lin et~al.}(2001)\citenamefont{Lin, Martinelli,
  Sachrajda, and Testa}}]{Lin:2001ek}
\bibinfo{author}{\bibfnamefont{C.~J.~D.} \bibnamefont{Lin}},
  \bibinfo{author}{\bibfnamefont{G.}~\bibnamefont{Martinelli}},
  \bibinfo{author}{\bibfnamefont{C.~T.} \bibnamefont{Sachrajda}},
  \bibnamefont{and} \bibinfo{author}{\bibfnamefont{M.}~\bibnamefont{Testa}},
  \bibinfo{journal}{Nucl. Phys.} \textbf{\bibinfo{volume}{B619}},
  \bibinfo{pages}{467} (\bibinfo{year}{2001}), \eprint{hep-lat/0104006}.

\bibitem[{\citenamefont{Aoki et~al.}(2005)}]{Aoki:2005uf}
\bibinfo{author}{\bibfnamefont{S.}~\bibnamefont{Aoki}} \bibnamefont{et~al.}
  (\bibinfo{collaboration}{CP-PACS}), \bibinfo{journal}{Phys. Rev.}
  \textbf{\bibinfo{volume}{D71}}, \bibinfo{pages}{094504}
  (\bibinfo{year}{2005}), \eprint{hep-lat/0503025}.

\bibitem[{\citenamefont{Ishizuka}(2009)}]{Ishizuka:2009bx}
\bibinfo{author}{\bibfnamefont{N.}~\bibnamefont{Ishizuka}},
  \bibinfo{journal}{PoS} \textbf{\bibinfo{volume}{LAT2009}},
  \bibinfo{pages}{119} (\bibinfo{year}{2009}), \eprint{0910.2772}.

\bibitem[{\citenamefont{Ishii et~al.}(2007)\citenamefont{Ishii, Aoki, and
  Hatsuda}}]{Ishii:2006ec}
\bibinfo{author}{\bibfnamefont{N.}~\bibnamefont{Ishii}},
  \bibinfo{author}{\bibfnamefont{S.}~\bibnamefont{Aoki}}, \bibnamefont{and}
  \bibinfo{author}{\bibfnamefont{T.}~\bibnamefont{Hatsuda}},
  \bibinfo{journal}{Phys. Rev. Lett.} \textbf{\bibinfo{volume}{99}},
  \bibinfo{pages}{022001} (\bibinfo{year}{2007}), \eprint{nucl-th/0611096}.

\bibitem[{\citenamefont{Aoki et~al.}(2010)\citenamefont{Aoki, Hatsuda, and
  Ishii}}]{Aoki:2009ji}
\bibinfo{author}{\bibfnamefont{S.}~\bibnamefont{Aoki}},
  \bibinfo{author}{\bibfnamefont{T.}~\bibnamefont{Hatsuda}}, \bibnamefont{and}
  \bibinfo{author}{\bibfnamefont{N.}~\bibnamefont{Ishii}},
  \bibinfo{journal}{Prog. Theor. Phys.} \textbf{\bibinfo{volume}{123}},
  \bibinfo{pages}{89} (\bibinfo{year}{2010}), \eprint{0909.5585}.

\bibitem[{\citenamefont{Aoki et~al.}(2012)\citenamefont{Aoki, Doi, Hatsuda,
  Ikeda, Inoue, Ishii, Murano, Nemura, and Sasaki}}]{Aoki:2012tk}
\bibinfo{author}{\bibfnamefont{S.}~\bibnamefont{Aoki}},
  \bibinfo{author}{\bibfnamefont{T.}~\bibnamefont{Doi}},
  \bibinfo{author}{\bibfnamefont{T.}~\bibnamefont{Hatsuda}},
  \bibinfo{author}{\bibfnamefont{Y.}~\bibnamefont{Ikeda}},
  \bibinfo{author}{\bibfnamefont{T.}~\bibnamefont{Inoue}},
  \bibinfo{author}{\bibfnamefont{N.}~\bibnamefont{Ishii}},
  \bibinfo{author}{\bibfnamefont{K.}~\bibnamefont{Murano}},
  \bibinfo{author}{\bibfnamefont{H.}~\bibnamefont{Nemura}}, \bibnamefont{and}
  \bibinfo{author}{\bibfnamefont{K.}~\bibnamefont{Sasaki}}
  (\bibinfo{collaboration}{HAL QCD}), \bibinfo{journal}{PTEP}
  \textbf{\bibinfo{volume}{2012}}, \bibinfo{pages}{01A105}
  (\bibinfo{year}{2012}), \eprint{1206.5088}.

\bibitem[{\citenamefont{Carbonell and Karmanov}(2016)}]{Carbonell:2016ekx}
\bibinfo{author}{\bibfnamefont{J.}~\bibnamefont{Carbonell}} \bibnamefont{and}
  \bibinfo{author}{\bibfnamefont{V.~A.} \bibnamefont{Karmanov}},
  \bibinfo{journal}{Phys. Lett.} \textbf{\bibinfo{volume}{B754}},
  \bibinfo{pages}{270} (\bibinfo{year}{2016}), \eprint{1601.00297}.

\bibitem[{\citenamefont{Aoki et~al.}(2013)\citenamefont{Aoki, Ishii, Doi,
  Ikeda, and Inoue}}]{Aoki:2013cra}
\bibinfo{author}{\bibfnamefont{S.}~\bibnamefont{Aoki}},
  \bibinfo{author}{\bibfnamefont{N.}~\bibnamefont{Ishii}},
  \bibinfo{author}{\bibfnamefont{T.}~\bibnamefont{Doi}},
  \bibinfo{author}{\bibfnamefont{Y.}~\bibnamefont{Ikeda}}, \bibnamefont{and}
  \bibinfo{author}{\bibfnamefont{T.}~\bibnamefont{Inoue}},
  \bibinfo{journal}{Phys. Rev.} \textbf{\bibinfo{volume}{D88}},
  \bibinfo{pages}{014036} (\bibinfo{year}{2013}), \eprint{1303.2210}.

\bibitem[{\citenamefont{Iritani et~al.}(2016)}]{Iritani:2016jie}
\bibinfo{author}{\bibfnamefont{T.}~\bibnamefont{Iritani}} \bibnamefont{et~al.},
  \bibinfo{journal}{JHEP} \textbf{\bibinfo{volume}{10}}, \bibinfo{pages}{101}
  (\bibinfo{year}{2016}), \eprint{1607.06371}.

\bibitem[{\citenamefont{Iritani et~al.}(2017)\citenamefont{Iritani, Aoki, Doi,
  Hatsuda, Ikeda, Inoue, Ishii, Nemura, and Sasaki}}]{Iritani:2017rlk}
\bibinfo{author}{\bibfnamefont{T.}~\bibnamefont{Iritani}},
  \bibinfo{author}{\bibfnamefont{S.}~\bibnamefont{Aoki}},
  \bibinfo{author}{\bibfnamefont{T.}~\bibnamefont{Doi}},
  \bibinfo{author}{\bibfnamefont{T.}~\bibnamefont{Hatsuda}},
  \bibinfo{author}{\bibfnamefont{Y.}~\bibnamefont{Ikeda}},
  \bibinfo{author}{\bibfnamefont{T.}~\bibnamefont{Inoue}},
  \bibinfo{author}{\bibfnamefont{N.}~\bibnamefont{Ishii}},
  \bibinfo{author}{\bibfnamefont{H.}~\bibnamefont{Nemura}}, \bibnamefont{and}
  \bibinfo{author}{\bibfnamefont{K.}~\bibnamefont{Sasaki}},
  \bibinfo{journal}{Phys. Rev.} \textbf{\bibinfo{volume}{D96}},
  \bibinfo{pages}{034521} (\bibinfo{year}{2017}), \eprint{1703.07210}.

\bibitem[{\citenamefont{Iritani et~al.}(2019)\citenamefont{Iritani, Aoki, Doi,
  Gongyo, Hatsuda, Ikeda, Inoue, Ishii, Nemura, and Sasaki}}]{Iritani:2018zbt}
\bibinfo{author}{\bibfnamefont{T.}~\bibnamefont{Iritani}},
  \bibinfo{author}{\bibfnamefont{S.}~\bibnamefont{Aoki}},
  \bibinfo{author}{\bibfnamefont{T.}~\bibnamefont{Doi}},
  \bibinfo{author}{\bibfnamefont{S.}~\bibnamefont{Gongyo}},
  \bibinfo{author}{\bibfnamefont{T.}~\bibnamefont{Hatsuda}},
  \bibinfo{author}{\bibfnamefont{Y.}~\bibnamefont{Ikeda}},
  \bibinfo{author}{\bibfnamefont{T.}~\bibnamefont{Inoue}},
  \bibinfo{author}{\bibfnamefont{N.}~\bibnamefont{Ishii}},
  \bibinfo{author}{\bibfnamefont{H.}~\bibnamefont{Nemura}}, \bibnamefont{and}
  \bibinfo{author}{\bibfnamefont{K.}~\bibnamefont{Sasaki}}
  (\bibinfo{collaboration}{HAL QCD}), \bibinfo{journal}{Phys. Rev.}
  \textbf{\bibinfo{volume}{D99}}, \bibinfo{pages}{014514}
  (\bibinfo{year}{2019}), \eprint{1805.02365}.

\bibitem[{\citenamefont{Nemura et~al.}(2009)\citenamefont{Nemura, Ishii, Aoki,
  and Hatsuda}}]{Nemura:2008sp}
\bibinfo{author}{\bibfnamefont{H.}~\bibnamefont{Nemura}},
  \bibinfo{author}{\bibfnamefont{N.}~\bibnamefont{Ishii}},
  \bibinfo{author}{\bibfnamefont{S.}~\bibnamefont{Aoki}}, \bibnamefont{and}
  \bibinfo{author}{\bibfnamefont{T.}~\bibnamefont{Hatsuda}},
  \bibinfo{journal}{Phys. Lett.} \textbf{\bibinfo{volume}{B673}},
  \bibinfo{pages}{136} (\bibinfo{year}{2009}), \eprint{0806.1094}.

\bibitem[{\citenamefont{Inoue et~al.}(2010)\citenamefont{Inoue, Ishii, Aoki,
  Doi, Hatsuda, Ikeda, Murano, Nemura, and Sasaki}}]{Inoue:2010hs}
\bibinfo{author}{\bibfnamefont{T.}~\bibnamefont{Inoue}},
  \bibinfo{author}{\bibfnamefont{N.}~\bibnamefont{Ishii}},
  \bibinfo{author}{\bibfnamefont{S.}~\bibnamefont{Aoki}},
  \bibinfo{author}{\bibfnamefont{T.}~\bibnamefont{Doi}},
  \bibinfo{author}{\bibfnamefont{T.}~\bibnamefont{Hatsuda}},
  \bibinfo{author}{\bibfnamefont{Y.}~\bibnamefont{Ikeda}},
  \bibinfo{author}{\bibfnamefont{K.}~\bibnamefont{Murano}},
  \bibinfo{author}{\bibfnamefont{H.}~\bibnamefont{Nemura}}, \bibnamefont{and}
  \bibinfo{author}{\bibfnamefont{K.}~\bibnamefont{Sasaki}}
  (\bibinfo{collaboration}{HAL QCD}), \bibinfo{journal}{Prog. Theor. Phys.}
  \textbf{\bibinfo{volume}{124}}, \bibinfo{pages}{591} (\bibinfo{year}{2010}),
  \eprint{1007.3559}.

\bibitem[{\citenamefont{Inoue et~al.}(2011)\citenamefont{Inoue, Ishii, Aoki,
  Doi, Hatsuda, Ikeda, Murano, Nemura, and Sasaki}}]{Inoue:2010es}
\bibinfo{author}{\bibfnamefont{T.}~\bibnamefont{Inoue}},
  \bibinfo{author}{\bibfnamefont{N.}~\bibnamefont{Ishii}},
  \bibinfo{author}{\bibfnamefont{S.}~\bibnamefont{Aoki}},
  \bibinfo{author}{\bibfnamefont{T.}~\bibnamefont{Doi}},
  \bibinfo{author}{\bibfnamefont{T.}~\bibnamefont{Hatsuda}},
  \bibinfo{author}{\bibfnamefont{Y.}~\bibnamefont{Ikeda}},
  \bibinfo{author}{\bibfnamefont{K.}~\bibnamefont{Murano}},
  \bibinfo{author}{\bibfnamefont{H.}~\bibnamefont{Nemura}}, \bibnamefont{and}
  \bibinfo{author}{\bibfnamefont{K.}~\bibnamefont{Sasaki}}
  (\bibinfo{collaboration}{HAL QCD}), \bibinfo{journal}{Phys. Rev. Lett.}
  \textbf{\bibinfo{volume}{106}}, \bibinfo{pages}{162002}
  (\bibinfo{year}{2011}), \eprint{1012.5928}.

\bibitem[{\citenamefont{Murano et~al.}(2011)\citenamefont{Murano, Ishii, Aoki,
  and Hatsuda}}]{Murano:2011nz}
\bibinfo{author}{\bibfnamefont{K.}~\bibnamefont{Murano}},
  \bibinfo{author}{\bibfnamefont{N.}~\bibnamefont{Ishii}},
  \bibinfo{author}{\bibfnamefont{S.}~\bibnamefont{Aoki}}, \bibnamefont{and}
  \bibinfo{author}{\bibfnamefont{T.}~\bibnamefont{Hatsuda}},
  \bibinfo{journal}{Prog. Theor. Phys.} \textbf{\bibinfo{volume}{125}},
  \bibinfo{pages}{1225} (\bibinfo{year}{2011}), \eprint{1103.0619}.

\bibitem[{\citenamefont{Doi et~al.}(2012)\citenamefont{Doi, Aoki, Hatsuda,
  Ikeda, Inoue, Ishii, Murano, Nemura, and Sasaki}}]{Doi:2011gq}
\bibinfo{author}{\bibfnamefont{T.}~\bibnamefont{Doi}},
  \bibinfo{author}{\bibfnamefont{S.}~\bibnamefont{Aoki}},
  \bibinfo{author}{\bibfnamefont{T.}~\bibnamefont{Hatsuda}},
  \bibinfo{author}{\bibfnamefont{Y.}~\bibnamefont{Ikeda}},
  \bibinfo{author}{\bibfnamefont{T.}~\bibnamefont{Inoue}},
  \bibinfo{author}{\bibfnamefont{N.}~\bibnamefont{Ishii}},
  \bibinfo{author}{\bibfnamefont{K.}~\bibnamefont{Murano}},
  \bibinfo{author}{\bibfnamefont{H.}~\bibnamefont{Nemura}}, \bibnamefont{and}
  \bibinfo{author}{\bibfnamefont{K.}~\bibnamefont{Sasaki}}
  (\bibinfo{collaboration}{HAL QCD}), \bibinfo{journal}{Prog. Theor. Phys.}
  \textbf{\bibinfo{volume}{127}}, \bibinfo{pages}{723} (\bibinfo{year}{2012}),
  \eprint{1106.2276}.

\bibitem[{\citenamefont{Inoue et~al.}(2012)\citenamefont{Inoue, Aoki, Doi,
  Hatsuda, Ikeda, Ishii, Murano, Nemura, and Sasaki}}]{Inoue:2011ai}
\bibinfo{author}{\bibfnamefont{T.}~\bibnamefont{Inoue}},
  \bibinfo{author}{\bibfnamefont{S.}~\bibnamefont{Aoki}},
  \bibinfo{author}{\bibfnamefont{T.}~\bibnamefont{Doi}},
  \bibinfo{author}{\bibfnamefont{T.}~\bibnamefont{Hatsuda}},
  \bibinfo{author}{\bibfnamefont{Y.}~\bibnamefont{Ikeda}},
  \bibinfo{author}{\bibfnamefont{N.}~\bibnamefont{Ishii}},
  \bibinfo{author}{\bibfnamefont{K.}~\bibnamefont{Murano}},
  \bibinfo{author}{\bibfnamefont{H.}~\bibnamefont{Nemura}}, \bibnamefont{and}
  \bibinfo{author}{\bibfnamefont{K.}~\bibnamefont{Sasaki}}
  (\bibinfo{collaboration}{HAL QCD}), \bibinfo{journal}{Nucl. Phys.}
  \textbf{\bibinfo{volume}{A881}}, \bibinfo{pages}{28} (\bibinfo{year}{2012}),
  \eprint{1112.5926}.

\bibitem[{\citenamefont{Murano et~al.}(2014)\citenamefont{Murano, Ishii, Aoki,
  Doi, Hatsuda, Ikeda, Inoue, Nemura, and Sasaki}}]{Murano:2013xxa}
\bibinfo{author}{\bibfnamefont{K.}~\bibnamefont{Murano}},
  \bibinfo{author}{\bibfnamefont{N.}~\bibnamefont{Ishii}},
  \bibinfo{author}{\bibfnamefont{S.}~\bibnamefont{Aoki}},
  \bibinfo{author}{\bibfnamefont{T.}~\bibnamefont{Doi}},
  \bibinfo{author}{\bibfnamefont{T.}~\bibnamefont{Hatsuda}},
  \bibinfo{author}{\bibfnamefont{Y.}~\bibnamefont{Ikeda}},
  \bibinfo{author}{\bibfnamefont{T.}~\bibnamefont{Inoue}},
  \bibinfo{author}{\bibfnamefont{H.}~\bibnamefont{Nemura}}, \bibnamefont{and}
  \bibinfo{author}{\bibfnamefont{K.}~\bibnamefont{Sasaki}}
  (\bibinfo{collaboration}{HAL QCD}), \bibinfo{journal}{Phys. Lett.}
  \textbf{\bibinfo{volume}{B735}}, \bibinfo{pages}{19} (\bibinfo{year}{2014}),
  \eprint{1305.2293}.

\bibitem[{\citenamefont{Kurth et~al.}(2013)\citenamefont{Kurth, Ishii, Doi,
  Aoki, and Hatsuda}}]{Kurth:2013tua}
\bibinfo{author}{\bibfnamefont{T.}~\bibnamefont{Kurth}},
  \bibinfo{author}{\bibfnamefont{N.}~\bibnamefont{Ishii}},
  \bibinfo{author}{\bibfnamefont{T.}~\bibnamefont{Doi}},
  \bibinfo{author}{\bibfnamefont{S.}~\bibnamefont{Aoki}}, \bibnamefont{and}
  \bibinfo{author}{\bibfnamefont{T.}~\bibnamefont{Hatsuda}},
  \bibinfo{journal}{JHEP} \textbf{\bibinfo{volume}{12}}, \bibinfo{pages}{015}
  (\bibinfo{year}{2013}), \eprint{1305.4462}.

\bibitem[{\citenamefont{Ikeda et~al.}(2014)\citenamefont{Ikeda, Charron, Aoki,
  Doi, Hatsuda, Inoue, Ishii, Murano, Nemura, and Sasaki}}]{Ikeda:2013vwa}
\bibinfo{author}{\bibfnamefont{Y.}~\bibnamefont{Ikeda}},
  \bibinfo{author}{\bibfnamefont{B.}~\bibnamefont{Charron}},
  \bibinfo{author}{\bibfnamefont{S.}~\bibnamefont{Aoki}},
  \bibinfo{author}{\bibfnamefont{T.}~\bibnamefont{Doi}},
  \bibinfo{author}{\bibfnamefont{T.}~\bibnamefont{Hatsuda}},
  \bibinfo{author}{\bibfnamefont{T.}~\bibnamefont{Inoue}},
  \bibinfo{author}{\bibfnamefont{N.}~\bibnamefont{Ishii}},
  \bibinfo{author}{\bibfnamefont{K.}~\bibnamefont{Murano}},
  \bibinfo{author}{\bibfnamefont{H.}~\bibnamefont{Nemura}}, \bibnamefont{and}
  \bibinfo{author}{\bibfnamefont{K.}~\bibnamefont{Sasaki}},
  \bibinfo{journal}{Phys. Lett.} \textbf{\bibinfo{volume}{B729}},
  \bibinfo{pages}{85} (\bibinfo{year}{2014}), \eprint{1311.6214}.

\bibitem[{\citenamefont{Etminan et~al.}(2014)\citenamefont{Etminan, Nemura,
  Aoki, Doi, Hatsuda, Ikeda, Inoue, Ishii, Murano, and
  Sasaki}}]{Etminan:2014tya}
\bibinfo{author}{\bibfnamefont{F.}~\bibnamefont{Etminan}},
  \bibinfo{author}{\bibfnamefont{H.}~\bibnamefont{Nemura}},
  \bibinfo{author}{\bibfnamefont{S.}~\bibnamefont{Aoki}},
  \bibinfo{author}{\bibfnamefont{T.}~\bibnamefont{Doi}},
  \bibinfo{author}{\bibfnamefont{T.}~\bibnamefont{Hatsuda}},
  \bibinfo{author}{\bibfnamefont{Y.}~\bibnamefont{Ikeda}},
  \bibinfo{author}{\bibfnamefont{T.}~\bibnamefont{Inoue}},
  \bibinfo{author}{\bibfnamefont{N.}~\bibnamefont{Ishii}},
  \bibinfo{author}{\bibfnamefont{K.}~\bibnamefont{Murano}}, \bibnamefont{and}
  \bibinfo{author}{\bibfnamefont{K.}~\bibnamefont{Sasaki}}
  (\bibinfo{collaboration}{HAL QCD}), \bibinfo{journal}{Nucl. Phys.}
  \textbf{\bibinfo{volume}{A928}}, \bibinfo{pages}{89} (\bibinfo{year}{2014}),
  \eprint{1403.7284}.

\bibitem[{\citenamefont{Yamada et~al.}(2015)\citenamefont{Yamada, Sasaki, Aoki,
  Doi, Hatsuda, Ikeda, Inoue, Ishii, Murano, and Nemura}}]{Yamada:2015cra}
\bibinfo{author}{\bibfnamefont{M.}~\bibnamefont{Yamada}},
  \bibinfo{author}{\bibfnamefont{K.}~\bibnamefont{Sasaki}},
  \bibinfo{author}{\bibfnamefont{S.}~\bibnamefont{Aoki}},
  \bibinfo{author}{\bibfnamefont{T.}~\bibnamefont{Doi}},
  \bibinfo{author}{\bibfnamefont{T.}~\bibnamefont{Hatsuda}},
  \bibinfo{author}{\bibfnamefont{Y.}~\bibnamefont{Ikeda}},
  \bibinfo{author}{\bibfnamefont{T.}~\bibnamefont{Inoue}},
  \bibinfo{author}{\bibfnamefont{N.}~\bibnamefont{Ishii}},
  \bibinfo{author}{\bibfnamefont{K.}~\bibnamefont{Murano}}, \bibnamefont{and}
  \bibinfo{author}{\bibfnamefont{H.}~\bibnamefont{Nemura}}
  (\bibinfo{collaboration}{HAL QCD}), \bibinfo{journal}{PTEP}
  \textbf{\bibinfo{volume}{2015}}, \bibinfo{pages}{071B01}
  (\bibinfo{year}{2015}), \eprint{1503.03189}.

\bibitem[{\citenamefont{Sasaki et~al.}(2015)\citenamefont{Sasaki, Aoki, Doi,
  Hatsuda, Ikeda, Inoue, Ishii, and Murano}}]{Sasaki:2015ifa}
\bibinfo{author}{\bibfnamefont{K.}~\bibnamefont{Sasaki}},
  \bibinfo{author}{\bibfnamefont{S.}~\bibnamefont{Aoki}},
  \bibinfo{author}{\bibfnamefont{T.}~\bibnamefont{Doi}},
  \bibinfo{author}{\bibfnamefont{T.}~\bibnamefont{Hatsuda}},
  \bibinfo{author}{\bibfnamefont{Y.}~\bibnamefont{Ikeda}},
  \bibinfo{author}{\bibfnamefont{T.}~\bibnamefont{Inoue}},
  \bibinfo{author}{\bibfnamefont{N.}~\bibnamefont{Ishii}}, \bibnamefont{and}
  \bibinfo{author}{\bibfnamefont{K.}~\bibnamefont{Murano}}
  (\bibinfo{collaboration}{HAL QCD}), \bibinfo{journal}{PTEP}
  \textbf{\bibinfo{volume}{2015}}, \bibinfo{pages}{113B01}
  (\bibinfo{year}{2015}), \eprint{1504.01717}.

\bibitem[{\citenamefont{Ikeda et~al.}(2016)\citenamefont{Ikeda, Aoki, Doi,
  Gongyo, Hatsuda, Inoue, Iritani, Ishii, Murano, and Sasaki}}]{Ikeda:2016zwx}
\bibinfo{author}{\bibfnamefont{Y.}~\bibnamefont{Ikeda}},
  \bibinfo{author}{\bibfnamefont{S.}~\bibnamefont{Aoki}},
  \bibinfo{author}{\bibfnamefont{T.}~\bibnamefont{Doi}},
  \bibinfo{author}{\bibfnamefont{S.}~\bibnamefont{Gongyo}},
  \bibinfo{author}{\bibfnamefont{T.}~\bibnamefont{Hatsuda}},
  \bibinfo{author}{\bibfnamefont{T.}~\bibnamefont{Inoue}},
  \bibinfo{author}{\bibfnamefont{T.}~\bibnamefont{Iritani}},
  \bibinfo{author}{\bibfnamefont{N.}~\bibnamefont{Ishii}},
  \bibinfo{author}{\bibfnamefont{K.}~\bibnamefont{Murano}}, \bibnamefont{and}
  \bibinfo{author}{\bibfnamefont{K.}~\bibnamefont{Sasaki}}
  (\bibinfo{collaboration}{HAL QCD}), \bibinfo{journal}{Phys. Rev. Lett.}
  \textbf{\bibinfo{volume}{117}}, \bibinfo{pages}{242001}
  (\bibinfo{year}{2016}), \eprint{1602.03465}.

\bibitem[{\citenamefont{Miyamoto et~al.}(2018)}]{Miyamoto:2017tjs}
\bibinfo{author}{\bibfnamefont{T.}~\bibnamefont{Miyamoto}}
  \bibnamefont{et~al.}, \bibinfo{journal}{Nucl. Phys.}
  \textbf{\bibinfo{volume}{A971}}, \bibinfo{pages}{113} (\bibinfo{year}{2018}),
  \eprint{1710.05545}.

\bibitem[{\citenamefont{Kawai et~al.}(2018)\citenamefont{Kawai, Aoki, Doi,
  Ikeda, Inoue, Iritani, Ishii, Miyamoto, Nemura, and Sasaki}}]{Kawai:2017goq}
\bibinfo{author}{\bibfnamefont{D.}~\bibnamefont{Kawai}},
  \bibinfo{author}{\bibfnamefont{S.}~\bibnamefont{Aoki}},
  \bibinfo{author}{\bibfnamefont{T.}~\bibnamefont{Doi}},
  \bibinfo{author}{\bibfnamefont{Y.}~\bibnamefont{Ikeda}},
  \bibinfo{author}{\bibfnamefont{T.}~\bibnamefont{Inoue}},
  \bibinfo{author}{\bibfnamefont{T.}~\bibnamefont{Iritani}},
  \bibinfo{author}{\bibfnamefont{N.}~\bibnamefont{Ishii}},
  \bibinfo{author}{\bibfnamefont{T.}~\bibnamefont{Miyamoto}},
  \bibinfo{author}{\bibfnamefont{H.}~\bibnamefont{Nemura}}, \bibnamefont{and}
  \bibinfo{author}{\bibfnamefont{K.}~\bibnamefont{Sasaki}}
  (\bibinfo{collaboration}{HAL QCD}), \bibinfo{journal}{PTEP}
  \textbf{\bibinfo{volume}{2018}}, \bibinfo{pages}{043B04}
  (\bibinfo{year}{2018}), \eprint{1711.01883}.

\bibitem[{\citenamefont{Ikeda}(2018)}]{Ikeda:2017mee}
\bibinfo{author}{\bibfnamefont{Y.}~\bibnamefont{Ikeda}}
  (\bibinfo{collaboration}{HAL QCD}), \bibinfo{journal}{J. Phys.}
  \textbf{\bibinfo{volume}{G45}}, \bibinfo{pages}{024002}
  (\bibinfo{year}{2018}), \eprint{1706.07300}.

\bibitem[{\citenamefont{Gongyo et~al.}(2018)}]{Gongyo:2017fjb}
\bibinfo{author}{\bibfnamefont{S.}~\bibnamefont{Gongyo}} \bibnamefont{et~al.},
  \bibinfo{journal}{Phys. Rev. Lett.} \textbf{\bibinfo{volume}{120}},
  \bibinfo{pages}{212001} (\bibinfo{year}{2018}), \eprint{1709.00654}.

\bibitem[{\citenamefont{Sasaki et~al.}(2017)}]{Sasaki:2017ysy}
\bibinfo{author}{\bibfnamefont{K.}~\bibnamefont{Sasaki}} \bibnamefont{et~al.},
  \bibinfo{journal}{PoS} \textbf{\bibinfo{volume}{LATTICE2016}},
  \bibinfo{pages}{116} (\bibinfo{year}{2017}), \eprint{1702.06241}.

\bibitem[{\citenamefont{Ishii et~al.}(2017)}]{Ishii:2017xud}
\bibinfo{author}{\bibfnamefont{N.}~\bibnamefont{Ishii}} \bibnamefont{et~al.},
  \bibinfo{journal}{PoS} \textbf{\bibinfo{volume}{LATTICE2016}},
  \bibinfo{pages}{127} (\bibinfo{year}{2017}), \eprint{1702.03495}.

\bibitem[{\citenamefont{Doi et~al.}(2017{\natexlab{a}})}]{Doi:2017cfx}
\bibinfo{author}{\bibfnamefont{T.}~\bibnamefont{Doi}} \bibnamefont{et~al.},
  \bibinfo{journal}{PoS} \textbf{\bibinfo{volume}{LATTICE2016}},
  \bibinfo{pages}{110} (\bibinfo{year}{2017}{\natexlab{a}}),
  \eprint{1702.01600}.

\bibitem[{\citenamefont{Nemura et~al.}(2017{\natexlab{a}})}]{Nemura:2017bbw}
\bibinfo{author}{\bibfnamefont{H.}~\bibnamefont{Nemura}} \bibnamefont{et~al.},
  \bibinfo{journal}{PoS} \textbf{\bibinfo{volume}{LATTICE2016}},
  \bibinfo{pages}{101} (\bibinfo{year}{2017}{\natexlab{a}}),
  \eprint{1702.00734}.

\bibitem[{\citenamefont{Doi et~al.}(2017{\natexlab{b}})}]{Doi:2017zov}
\bibinfo{author}{\bibfnamefont{T.}~\bibnamefont{Doi}} \bibnamefont{et~al.}, in
  \emph{\bibinfo{booktitle}{{35th International Symposium on Lattice Field
  Theory (Lattice 2017) Granada, Spain, June 18-24, 2017}}}
  (\bibinfo{year}{2017}{\natexlab{b}}), \eprint{1711.01952},
  \urlprefix\url{http://inspirehep.net/record/1634621/files/arXiv:1711.01952.pdf}.

\bibitem[{\citenamefont{Nemura et~al.}(2017{\natexlab{b}})}]{Nemura:2017vjc}
\bibinfo{author}{\bibfnamefont{H.}~\bibnamefont{Nemura}} \bibnamefont{et~al.},
  in \emph{\bibinfo{booktitle}{{35th International Symposium on Lattice Field
  Theory (Lattice 2017) Granada, Spain, June 18-24, 2017}}}
  (\bibinfo{year}{2017}{\natexlab{b}}), \eprint{1711.07003},
  \urlprefix\url{http://inspirehep.net/record/1637203/files/arXiv:1711.07003.pdf}.

\bibitem[{\citenamefont{Gongyo and Aoki}(2018)}]{Gongyo:2018gou}
\bibinfo{author}{\bibfnamefont{S.}~\bibnamefont{Gongyo}} \bibnamefont{and}
  \bibinfo{author}{\bibfnamefont{S.}~\bibnamefont{Aoki}},
  \bibinfo{journal}{PTEP} \textbf{\bibinfo{volume}{2018}},
  \bibinfo{pages}{093B03} (\bibinfo{year}{2018}), \eprint{1807.02967}.

\bibitem[{\citenamefont{Yamazaki et~al.}(2012)\citenamefont{Yamazaki, Ishikawa,
  Kuramashi, and Ukawa}}]{Yamazaki:2012hi}
\bibinfo{author}{\bibfnamefont{T.}~\bibnamefont{Yamazaki}},
  \bibinfo{author}{\bibfnamefont{K.-i.} \bibnamefont{Ishikawa}},
  \bibinfo{author}{\bibfnamefont{Y.}~\bibnamefont{Kuramashi}},
  \bibnamefont{and} \bibinfo{author}{\bibfnamefont{A.}~\bibnamefont{Ukawa}},
  \bibinfo{journal}{Phys. Rev.} \textbf{\bibinfo{volume}{D86}},
  \bibinfo{pages}{074514} (\bibinfo{year}{2012}), \eprint{1207.4277}.

\end{thebibliography}
\end{document}